\documentclass[sn-standardnature]{sn-jnl}

\usepackage{subfig} 
\usepackage{doi}
\usepackage{pdflscape}	
\usepackage{lineno}
\usepackage[none]{hyphenat}
\usepackage{caption}
\usepackage{subcaption}
\usepackage{graphicx}
\usepackage{amsmath}
\usepackage{array}

\usepackage{tabularx}
\usepackage{gensymb}

\newcommand{\dmunits}{\ensuremath{{\rm pc \, cm^{-3}}}}
\newcommand{\Xrayunits}{\ensuremath{{\rm erg \, cm^{-2} \, s^{-1}}}}
\newcommand{\luminu}{\ensuremath{{\rm erg \, s^{-1}}}}
\newcommand{\rmunits}{\ensuremath{{\rm rad \, m^{-2}}}}
\newcommand{\pdotunits}{\ensuremath{{\rm s \, s^{-1}}}}

\newcommand{\ulpo}{ASKAP\,J1839\ensuremath{-}0756}

\newcommand{\aj}{Astron. J.}   
\newcommand{\apj}{Astrophys. J.}   
\newcommand{\apjl}{Astrophys. J. Lett.}   
\newcommand{\apjs}{Astrophys. J. Suppl. Ser.}   
\newcommand{\aap}{Astron. Astrophys.}   
\newcommand{\aaps}{Astron. Astrophys. Suppl.}   
\newcommand{\mnras}{Mon. Not. R. Astron. Soc.}   
\newcommand{\nat}{Nature} 
\newcommand{\prd}{Phys. Rev. D}   
\newcommand{\pasa}{Publ. Astron. Soc. Aust.}   
\newcommand{\pasp}{Publ. Astron. Soc. Pac.}   

\begin{document}
\title
{A 6.45-hour period coherent radio transient emitting interpulses}

\author*[1,2,3]{\fnm{Y.~W.~J.} \sur{Lee}}\email{ylee2156@uni.sydney.edu.au}
\author*[1,2]{\fnm{M.} \sur{Caleb}}\email{manisha.caleb@sydney.edu.au}
\author[1,2]{\fnm{Tara} \sur{Murphy}}
\author[3]{\fnm{E.} \sur{Lenc}}
\author[4]{\fnm{D.~L.} \sur{Kaplan}}
\author[5]{{\fnm{L.} \sur{Ferrario}}}
\author[6,7,8]{{\fnm{Z.} \sur{Wadisingh}}}
\author[4]{{\fnm{A.} \sur{Anumarlapudi}}}
\author[9]{{\fnm{N.} \sur{Hurley-Walker}}}
\author[10]{{\fnm{V.} \sur{Karambelkar}}}
\author[10,11]{{\fnm{S.~K.} \sur{Ocker}}}
\author[9]{{\fnm{S.} \sur{McSweeney}}}
\author[12]{{\fnm{H.} \sur{Qiu}}}
\author[13]{{\fnm{K.~M.} \sur{Rajwade}}}
\author[3]{{\fnm{A.} \sur{Zic}}}
\author[3]{{\fnm{K.~W.} \sur{Bannister}}}
\author[9]{{\fnm{N.~D.~R.} \sur{Bhat}}} 
\author[2,14]{{\fnm{A.} \sur{Deller}}}
\author[1,2]{{\fnm{D.} \sur{Dobie}}}
\author[1]{{\fnm{L.} \sur{Driessen}}}
\author[7]{{\fnm{K.} \sur{Gendreau}}}
\author[9]{{\fnm{M.} \sur{Glowacki}}}
\author[3]{{\fnm{V.} \sur{Gupta}}}
\author[14]{{\fnm{J.~N.} \sur{Jahns-Schindler}}}
\author[14]{{\fnm{A.} \sur{Jaini}}}
\author[9]{{\fnm{C.~W.} \sur{James}}}
\author[10]{{\fnm{M.~M.} \sur{Kasliwal}}}
\author[3]{{\fnm{M.~E.} \sur{Lower}}}
\author[14]{{\fnm{R.~M.} \sur{Shannon}}}
\author[14]{{\fnm{P.~A.} \sur{Uttarkar}}}
\author[14]{{\fnm{Y.} \sur{Wang}}}
\author[9]{{\fnm{Z.} \sur{Wang}}}

\affil[1]{\orgdiv{Sydney Institute for Astronomy, School of Physics}, \orgname{The University of Sydney}, \orgaddress{\city{Sydney}, \postcode{2006}, \state{NSW}, \country{Australia}}}

\affil[2]{\orgname{ARC Centre of Excellence for Gravitational Wave Discovery (OzGrav)}, \orgaddress{ \city{Hawthorn}, \postcode{3122}, \state{Victoria}, \country{Australia}}}

\affil[3]{\orgname{Australia Telescope National Facility, CSIRO, Space \& Astronomy}, \orgaddress{\street{PO Box 76}, \city{Epping}, \postcode{1710}, \state{NSW}, \country{Australia}}}

\affil[4]{\orgname{Center for Gravitation, Cosmology, and Astrophysics, Department of Physics, University of Wisconsin-Milwaukee}, \orgaddress{\street{P.O. Box 413}, \city{Milwaukee}, \postcode{53201}, \state{WI}, \country{USA}}}

\affil[5]{\orgname{Mathematical Sciences Institute, The Australian National University}, \orgaddress{ \city{Canberra}, \state{ACT}, \postcode{2601}, \country{Australia}}}

\affil[6]{\orgname{Department of Astronomy, University of Maryland}, \orgaddress{College Park}, \postcode{MD 20742-4111}, \country{USA}}

\affil[7]{\orgdiv{Astrophysics Science Division}, \orgname{NASA Goddard Space Flight Center}, \orgaddress{\street{8800 Greenbelt Road}, \city{Greenbelt}, \state{MD}, \postcode{20771}, \country{USA}}}

\affil[8]{\orgname{Center for Research and Exploration in Space Science and Technology, NASA/GSFC}, \orgaddress{Greenbelt}, \postcode{MD 20771}, \country{USA}}

\affil[9]{\orgname{International Centre for Radio Astronomy Research, Curtin University}, \orgaddress{\street{Kent Street}, \city{Bentley WA}, \postcode{6102}, \country{Australia}}}

\affil[10]{\orgname{Cahill Center for Astronomy and Astrophysics, California Institute of Technology}, \orgaddress{\city{Pasadena}, \postcode{CA 91125}, \country{USA}}}

\affil[11]{\orgname{Carnegie Science Observatories}, \orgaddress{\city{Pasadena}, \postcode{CA 91101}, \country{USA}}}

\affil[12]{\orgname{SKA Observatory, Jodrell Bank}, \orgaddress{ \city{Lower Withington, Macclesfield}, \state{Cheshire}, \postcode{SK11 9FT}, \country{UK}}}
\affil[13]{\orgname{Astrophysics, University of Oxford, Denys Wilkinson Building}, \orgaddress{Keble Road, Oxford}, \postcode{OX1 3RH}, \country{UK}}

\affil[14]{\orgname{Centre for Astrophysics and Supercomputing, Swinburne University of Technology}, \orgaddress{\street{John Street}, \city{Hawthorn}, \postcode{3122}, \country{Australia}}}

\abstract{
Long-period radio transients are a novel class of astronomical objects characterised by prolonged periods ranging from 18 minutes to 54 minutes. They exhibit highly polarised, coherent, beamed radio emission lasting only 10--100 seconds. The intrinsic nature of these objects is subject to speculation, with highly magnetised white dwarfs and neutron stars being the prevailing candidates. Here we present ASKAP J183950.5$-$075635.0 (hereafter, \ulpo{}), boasting the longest known period of this class at 6.45 hours. It exhibits emission characteristics of an ordered dipolar magnetic field, with pulsar-like bright main pulses and weaker interpulses offset by about half a period are indicative of an oblique or orthogonal rotator. This phenomenon, observed for the first time in a long-period radio transient, confirms that the radio emission originates from both magnetic poles and that the observed period corresponds to the rotation period. The spectroscopic and polarimetric properties of \ulpo{} are consistent with a neutron star origin, and this object is a crucial piece of evidence in our understanding of long-period radio sources and their links to neutron stars.
}

\maketitle
\ulpo{} was discovered with the Australian SKA Pathfinder (ASKAP) telescope in a 15-minute Rapid ASKAP Continuum Survey low-band (RACS-low) at 943.5\,MHz \citep{RacsLow-Duchesne, RacsLow-McConnell} observation on 2024-01-26. The source was deemed noteworthy due to its point source nature, substantial circular polarisation fraction and absence of a known object with continuum radio emission in archival data. The light curve of the detection at 10-second time resolution showed the flux density rapidly decaying from 0.70\,Jy to 0.03\,Jy over just 15 minutes. The burst was highly polarised with a linear polarisation fraction of 90\% and a circular polarisation fraction of 37\%. Simultaneous time-series data at 13.8\,ms resolution showed several narrow sub-pulses superimposed on the burst profile. 
The burst's rotation measure (RM) of $214 \pm 1.3 \, \rmunits$ is consistent with that of the smoothed Galactic foreground and pulsars near the source \citep{ATNFCatalogue}.  This consistency precludes the presence of a substantial intrinsic RM imparted close to the source.

\begin{figure}
    \includegraphics[width=0.85\linewidth]{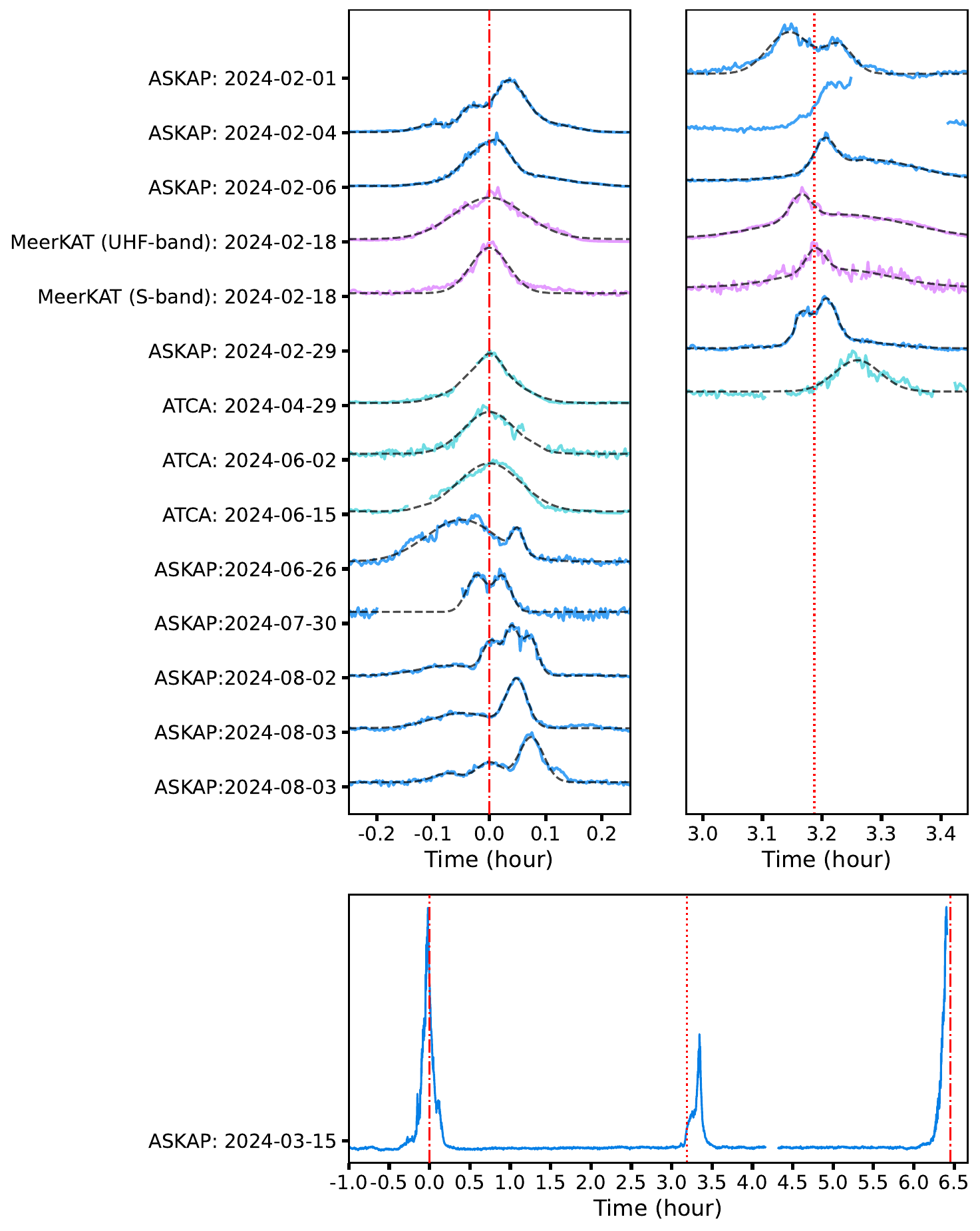}
    \caption{Light curves of follow-up observations on \ulpo{}. The upper left panel shows the main pulses aligned at the time of arrival (ToA), while the upper right panel shows the weaker secondary pulses seen in the longer observations roughly 3.2 hours after the main pulse. Each pulse is normalised individually to enhance readability. The different colours represent the telescope used, and the black dashed lines show the pulse profiles used to determine each pulse ToA (Methods). The flux densities of the interpulses are typically 10--20\% of that of the main pulses. The bottom panel displays the full light curve of the ASKAP observation on 2024-03-15, showing a period of 6.4\,hours. The red dot-dashed lines represent the expected ToA of the main pulse, while the red dotted line indicates the best-fitted phase difference of the weaker pulses (Methods). The arrival times of the weaker pulses fluctuate slightly, a phenomenon also observed in radio pulsars, causing them to deviate from the best-fit line.}
    \label{fig:light curve}
\end{figure}

Following the discovery, we conducted 14 radio follow-up observations between 2024-02-01 and 2024-08-03: ten with ASKAP (800--1088 MHz), one in dual frequency sub-array mode with the MeerKAT radio interferometer (UHF-band: 544--1088 MHz and S-band: 1750--2625 MHz), and three with the Australia Telescope Compact Array (ATCA) (1076--3124 MHz). Table \ref{tab:ObservationDetails} summarises the details of, and detection(s) in, each of these observations. Overall, \ulpo{} was consistently detected with at least one pulse in each epoch, with peak flux densities ranging from 0.01\,Jy to 1.4\,Jy. The MeerKAT observation allowed us to refine our initial ASKAP localisation of \ulpo{} to RA (J2000) = $18^h39^m50.57^s \pm 0.33''$ and Dec (J2000) = $-07^d56'39.17'' \pm 0.15''$ with 1-$\sigma$ uncertainty, which places it $1\degree$ below the Galactic plane. By de-dispersing a sub-pulse found in the MeerKAT observation, we estimated the dispersion measure (DM) to be $188.4 \pm 1.7 \, \dmunits$ at 1-$\sigma$ error (see Methods and Extended Data Figure \ref{fig:subpulse}). Based on the Galactic electron density models of NE2001 \citep{NE2001} and YMW16 \citep{YMW16}, we inferred \ulpo{} to be $4.0\pm 1.2$\,kpc away. The ASKAP observation on 2024-03-15 contained two bright pulses separated by 6.425 hours, providing a preliminary estimate of the source period. Figure \ref{fig:light curve} shows the light curves of all follow-up observations. 

In observations between 2024-02-03 and 2024-04-29, we noticed a weaker secondary pulse at approximately 3.2 hours after the brighter main pulse (see Figure \ref{fig:light curve}). These weaker pulses appearing at roughly half the period (i.e., $180\degree$ longitude separation from the main pulse, where $360\degree$ represents the whole period) are reminiscent of interpulses commonly seen in radio pulsars \citep{InterpulsePulsarCatalogue}. Interpulses are generally thought to indicate that the observer is sampling antipodal magnetic poles of a compact object. For collimated emission, this may occur when the magnetic axis is almost orthogonal to the rotation axis of the star. However, we note that no interpulse was detected in ASKAP observations starting from 2024-07-30, despite the observation duration encompassing the period of the source. This absence can be attributed to the flux density dropping below the sensitivity of the telescope or an intrinsic change in the source itself. 

We used the times of arrival (ToAs) of 21 pulses and interpulses (seen in Figure \ref{fig:light curve}) in \texttt{PINT} \citep{PINT}, a standard pulsar timing tool, to estimate a period. The interpulse on 2024-02-04 was omitted as the observation was interrupted. We estimated a period at 1-$\sigma$ error to be $23221.7 \pm 0.1$ seconds (6.45\,hours), establishing \ulpo{} as the longest-period coherent radio transient discovered to date (see Methods). The 3$\sigma$ limit on the period derivative is $\lvert\dot{P}\rvert <1.6\times 10^{-7}\,{\rm s\,s}^{-1}$, implying an upper limit on the spin-down luminosity of $\dot{E} \lesssim 10^{26} \luminu$. Table \ref{tab:sourceparams} lists the observed and derived parameters of \ulpo{}. We note that \ulpo{} has a period similar to 1E~161348$-$5055, a young magnetar with a period of 6.67 hours, embedded in a supernova remnant \citep{1E1613_Discovery, 1E1613_Period}. However, 1E~161348$-$5055 is apparently radio-quiet and emits mostly in the X-ray part of the spectrum with a potential infrared counterpart \citep{1E1613_IRcounterpart}. 

\begin{landscape}
    \begin{figure}
    \centering
    \includegraphics[width=1\linewidth]{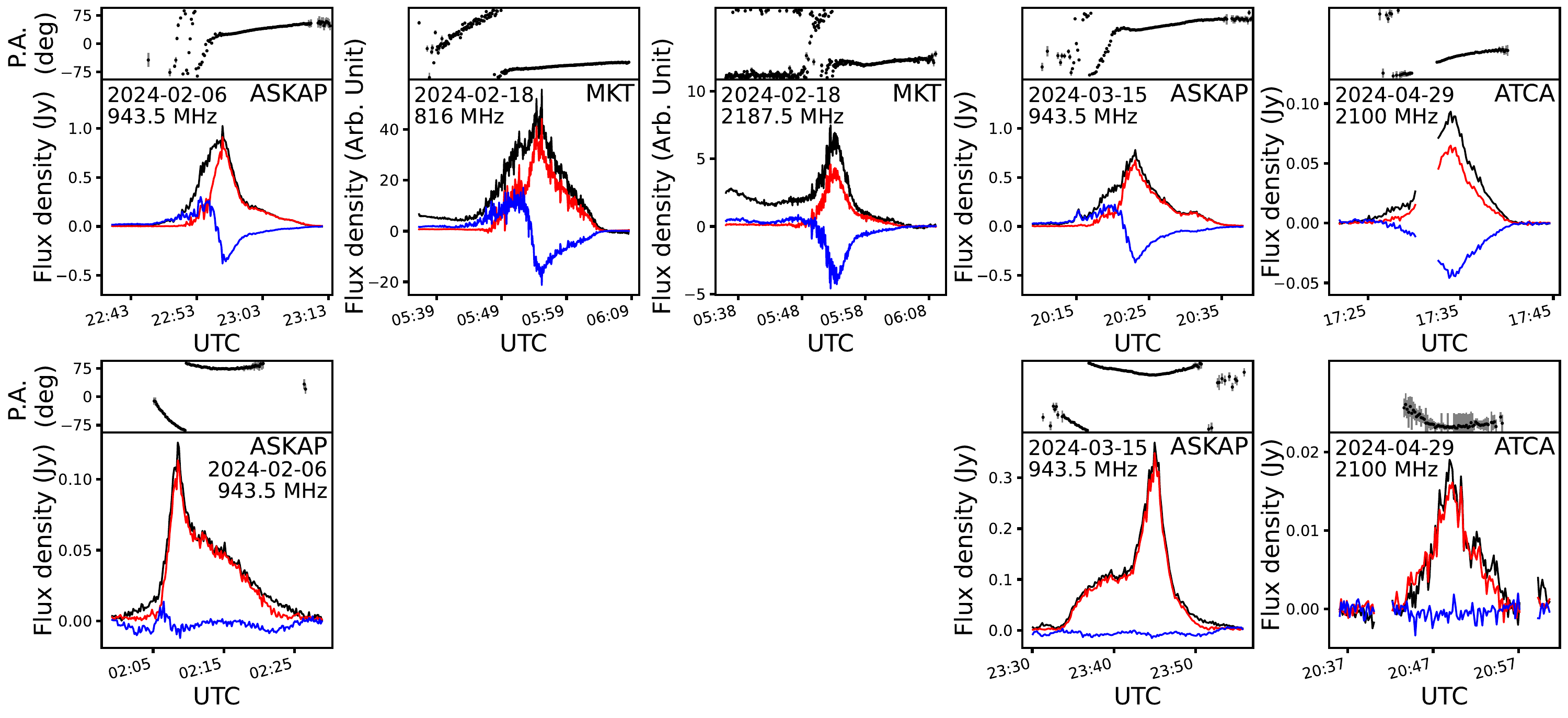}
    \caption{Pulse profiles from selected observations with ASKAP, MeerKAT (MKT), and ATCA. The top row displays the main pulses while the bottom row shows the interpulses. In each subplot, the upper panel shows the polarisation position angle (PA) swing of the pulse in degrees and the lower panel shows the total intensity (black), the linear polarisation intensity (red), and the circular polarisation intensity (blue). The distinct features in the PA swing (i.e., positive slope in main pulses and U-shaped interpulses) and the circular polarisation (i.e., strong circular polarisation with sign change in the main pulse) allowed us to distinguish and identify interpulses (bottom row) from the main pulses (top row) before the source's period was well-determined. The unit of the MeerKAT data is arbitrary since MeerKAT beamformed data are not flux calibrated. The interpulses of the MeerKAT observation are not shown due to poor data quality and strong radio frequency interference.}
    \label{fig:pulseprofile}
    \end{figure}
\end{landscape}

There are three reasons to suggest that the weaker secondary pulses observed in \ulpo{} in Figure \ref{fig:light curve} are interpulses associated with the brighter main pulses:

\begin{enumerate}
    \item The weaker secondary pulses have an $177.8\degree \pm 1.0 \degree$ (1-$\sigma$ error) longitude difference from the bright main pulses;
    \item The slope of the polarisation position angle (PA) of the weaker secondary pulse is reversed compared to their brighter counterparts, a characteristic observed in pulsars exhibiting interpulses \citep{Johnston_Props_of_IP} (Figure \ref{fig:pulseprofile});
    \item The fractional circular polarisation of the weaker secondary pulse is consistently different to that of the brighter main pulse in all observations (Figure \ref{fig:pulseprofile}).
\end{enumerate} 

\noindent Based on the above, we concluded that the secondary pulse is a {\em bona-fide} interpulse and not a weaker main pulse due to fluctuations in flux density, and that \ulpo{} is the first long-period radio transient to show interpulse emission. Consequently, we established that the observed period is a rotational period. 

The durations of the main pulses in the ASKAP observations were typically  between 320--710 seconds, which corresponds to a duty cycle of 1.4 -- 3.1\%. This duty cycle is broadly consistent with those seen in other long-period transients \citep{GPM1839, GLEAM-X1627, ASKAPJ1935}, and is also comparable to that of slow pulsars (5\,s $< P <$15\,s), which have duty cycles of $0.01 \% - 6 \%$ \citep{ATNFCatalogue}. Using a distance of 4.0\,kpc, we estimated the radio luminosity of the source in February 2024 (i.e., when it was first discovered) to be $L \approx 4 \times 10^{31} (\Omega_{1\rm GHz})\,\luminu \gtrsim 2 \times 10^{28}\,\luminu$, where $\Omega_{1\rm GHz}$ is the beam solid angle (see Methods). A caveat is that this beaming calculation is based on scaling relations of ordinary radio pulsars.  On average, the main pulses of \ulpo{} are 80\% linearly polarised and 40\% circularly polarised, while the interpulses are 90\% linearly polarised with negligible circular polarisation. In Figure \ref{fig:pulseprofile}, we also notice distinct features in the PA swing, where main pulses have a positive slope and interpulses are U-shaped. Overall, we observe similarities between \ulpo{} and the radio pulsar PSR~B1702$–$19, where the main pulses exhibit strong circular polarisation, the interpulses are nearly 100\% linearly polarized, and the interpulse PA swing follows a U-shaped pattern \citep{PSRB1702-19}. 

\begin{figure}
    \centering
    \includegraphics[width=0.8\linewidth]{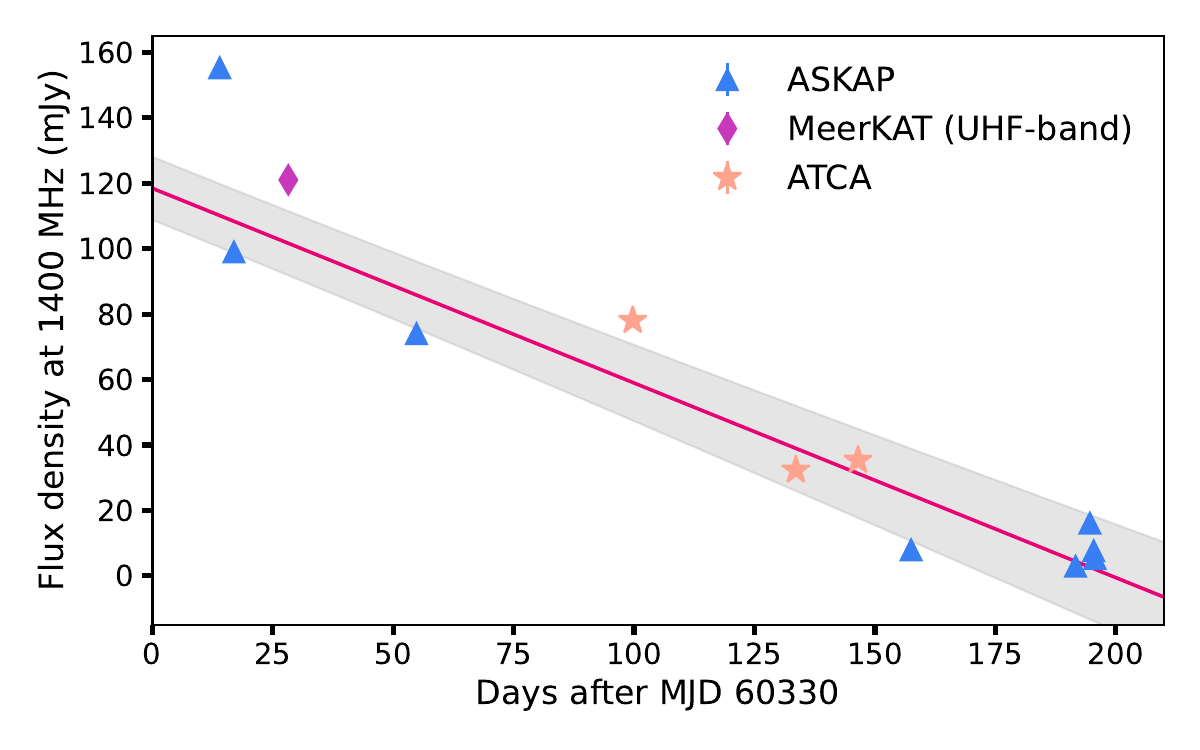}
    \caption{Main pulse flux density scaled to 1400 MHz as a function of time. The flux density of all observed main pulses are scaled to 1.4\,GHz according to their best-fit spectral index (Methods). The line represents a linear fit to the data with a slope of $-0.60 \pm 0.07 \, \rm mJy\, day^{-1}$, with the shaded region indicating the 1-$\sigma$ error. The statistical uncertainties in the flux density are approximately 1--3\%, making the error bars too short to be visible in the plot.}
    \label{fig:decreasinglumin}
\end{figure}

The RMs measured from all follow-up ASKAP observations were consistently between $214 \pm 1.0$ and $221 \pm 1.9$\,\rmunits, suggesting that the magnetic field along the line of sight did not vary notably in this time. We see phase-resolved RM variability in the ASKAP and MeerKAT data, which can be a sign of wave mode coupling or Faraday Conversion (see Methods and Extended Data Figure \ref{fig:rmvar} and \ref{fig:rm_v_var}). By imaging the MeerKAT UHF-band data between the main pulse and interpulse, we found no off-pulse emission down to an RMS intensity of 90\,$\mu \rm Jy$, indicating that no pulsar wind nebula or extended emission surrounds \ulpo{} (see Extended Data Figure \ref{fig:MeerKAT_Images}). The spectral index ($\alpha$) of the main pulse varied between observations ($-2.99\leq \alpha \leq -2.27$). This is much steeper compared to the mean spectral index of radio pulsars ($\alpha = -1.60 \pm 0.03$ \citep{Pulsar_Spectral_Index}) but still within the range of radio pulsars (e.g., PSR B1844$-$04 ($\alpha < -2.64$) and PSR B1919$+$21 ($\alpha < -2.88$) \citep{SteepSpectrumPulsar}) and comparable to some radio-loud magnetars (e.g., Swift J1818.0$-$1607 ($\alpha \sim -2.0$) \citep{10.1093/mnras/staa3789}). By scaling the flux densities of the main pulses (see Table \ref{tab:ObservationDetails}) to 1400 MHz, we noticed a secular decline in the flux densities with time (Figure \ref{fig:decreasinglumin}). This decline is consistent with the decreasing radio luminosities observed in the six currently known radio-loud magnetars \citep{MagnetarCatalogue}. The Murchison Widefield Array (MWA; 170--200\,MHz) was on source simultaneously with ASKAP on 2024-08-02 during the on-pulse period. Correcting for dispersion delay, no pulse was detected with the MWA down to a (3-$\sigma$) limit of 60\,mJy\,beam$^{-1}$; extrapolating from the flux density and spectral index of the ASKAP pulse observed at that time, a pulse flux density of $S_\mathrm{185MHz}\sim2.3$\,Jy would be expected. We thus concluded that the spectrum turns over between $\sim$300--600\,MHz (see Extended Data Figure \ref{fig:radio_SED}), a feature previously unseen in such objects. 

Compared to other long-period radio transients, \ulpo{} has a period $\sim 7$ times longer than ASKAP J1935$+$2148 \citep{ASKAPJ1935} and $\sim 20$ times longer than GLEAM-X J162759.5$-$523504.3202 \citep{GLEAM-X1627} and GPM J1839$-$10 \citep{GPM1839}. However, in terms of the other properties, \ulpo{} is similar to other long-period radio transients. Overall, the duty cycles of these sources are typically $< 10 \%$, with GPM J1839$-$10 being the only exception, occasionally emitting wide pulses with a duty cycle exceeding 20 \%. The similar duty cycles suggest a comparable beam size across this class of objects. Additionally, all the sources possess radio luminosities greater than their inferred spin-down luminosities, indicating that the radio emission is not rotationally powered. Furthermore, GLEAM-X J162759.5$-$523504.3202 was active only for three months, and ASKAP J1935$+$2148 was only detectable for eight months. Given the decreasing trend in radio luminosity of \ulpo{} (Figure \ref{fig:decreasinglumin}), it may well have a similar active lifetime to these sources. GLEAM-X J162759.5$-$523504.3202 exhibited two active periods that were separated by a null interval in its active lifetime. Ongoing monitoring on \ulpo{} may observe a rebrightening, which can provide a stronger connection between these two sources. Altogether, these similarities suggest that all long-period radio transients potentially share a common origin and are likely magnetically powered. 

\begin{figure}
    \centering
    \includegraphics[width=1\linewidth]{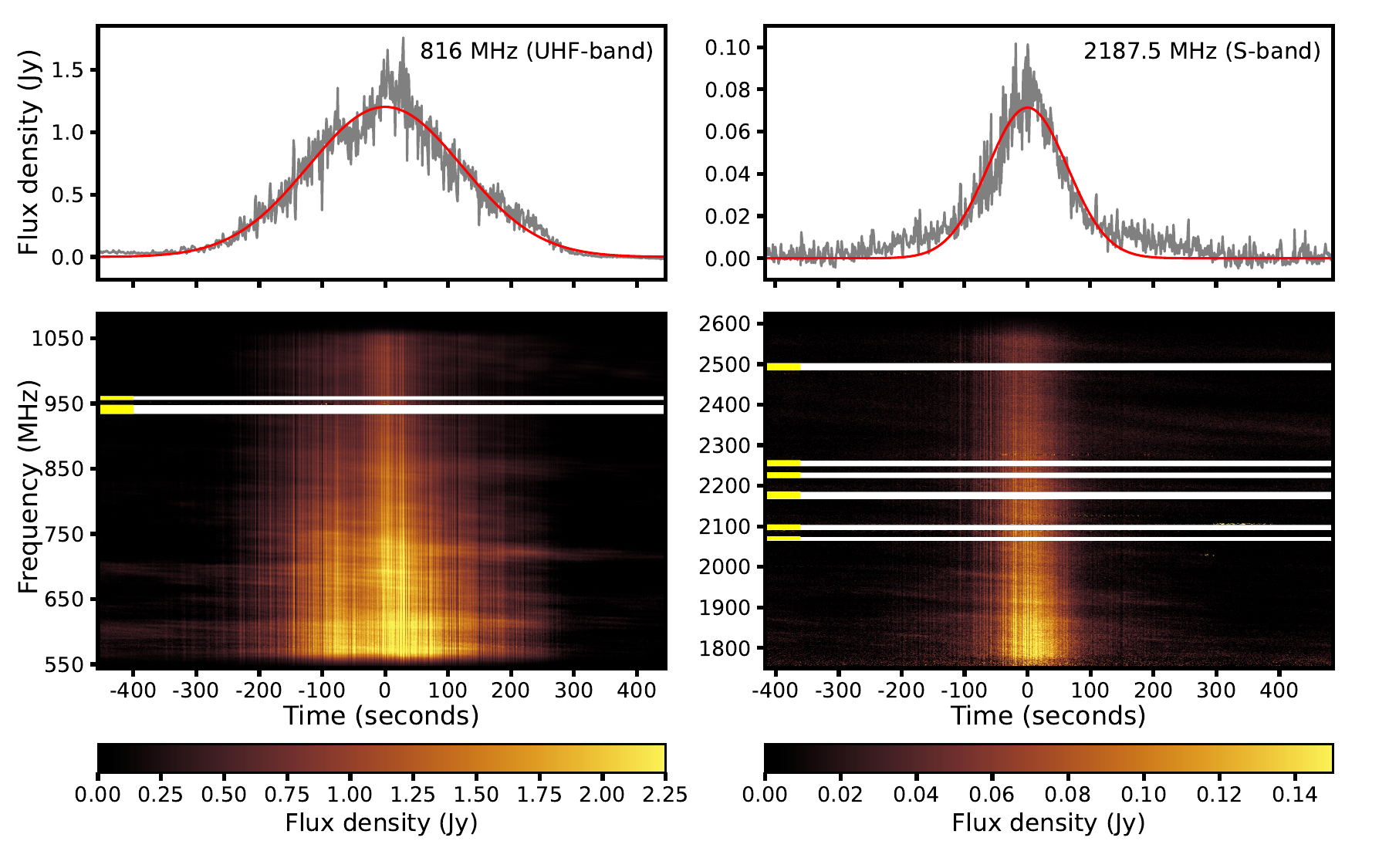}
    \caption{Light curves and dynamic spectra of the MeerKAT detection at UHF-band and S-band. In each plot, the upper and lower panel display the frequency-averaged light curve and the dynamic spectrum, respectively. A Gaussian curve (red) is fitted to each light curve. We found that the pulse width in the UHF-band is nearly twice as wide as that in the S-band, providing strong evidence for radius-to-frequency mapping. Quasi-periodic microstructure on (sub-)second scales can also be seen in both dynamic spectra. Horizontal white and yellow stripes represent the flagged radio frequency interference (RFI). The horizontal features at lower frequencies (550\,MHz -- 850\,MHz) are RFI that were unable to be removed without off-source beam baseline subtraction. These features were also seen in the off-pulse interval, so we conclude that they are not genuine astrophysical features related to the source.}
    \label{fig:MeerKAT_pulse}
\end{figure}

The MeerKAT observation on 2024-02-18 was conducted in dual frequency subarray mode, which allowed the detection of a main pulse and an interpulse over a wide and continuous range of frequencies. By fitting a Gaussian curve to the main pulse, we measured the full-width at half the maximum in the UHF-band (544–1088 MHz) to be 286.4 seconds. This width is almost twice as wide as that of the same pulse detected in the S-band (1750-2625 MHz), which is measured to be 148.4 seconds (Figure \ref{fig:MeerKAT_pulse}). This finding provides strong evidence for radius-to-frequency mapping in \ulpo{}, suggesting that the emission height decreases with increasing frequency within the beam \citep{RS75}. The MeerKAT detections also exhibited strong, quasi-periodic fine structure as seen in Extended Data Figure \ref{fig:MeerKAT_Stokes_Uband} and \ref{fig:MeerKAT_Stokes_Sband}, with features as narrow as 100\,ms. Quasi-periodicities typically appear as repeating micropulses \citep{bor76} that are part of a microstructure superimposed on the wider sub-pulses (e.g.\,\citep{lgs12}). Microstructure is usually theorised to be caused by mechanisms related to magnetospheric radio emission or its propagation through the magnetosphere, and there is often a variety of timescales observed even within a given source \citep[e.g.][]{Lange1998}. 
An auto-correlation function analysis revealed the most common quasi-period seen in \ulpo{} to be 2.4 seconds. This value is an order of magnitude smaller than the empirical scaling between quasi-period and spin-period seen in normal pulsars ($P_{\mu} \approx 10^{-3} P_{\rm spin}$) \citep{PulsarQuasi-periodicity} and magnetars. It is possible that long-period radio transients exhibit a different relationship compared to typical pulsars and magnetars, possibly due to differences in emission mechanisms and/or evolutionary stages. More sources are needed to generalize and unify the properties of radio pulsars, magnetars and long-period radio transients. By measuring the width and flux density of the microstructure, we estimated the peak brightness temperature of \ulpo{} to be $\sim 10^{20}\,\rm K$.

Periodic radio emissions on a minute timescale have been detected from white dwarf binaries with an M-dwarf companion, which emits in the optical and near infrared spectrum \citep{ARSco,J1912}. To search for a possible main-sequence companion star in \ulpo{}, we conducted deep near-infrared observations using the Wide Field Infrared Camera (WIRC) \citep{WIRC} on the Palomar Observatory and the FourStar camera \citep{FourStar} on the Magellan Baade 6.5-m telescope. The resulting images are shown in Extended Data Figure \ref{fig:OIR}. No source was detected within the radio error ellipse.  Based on the statistics of the nearby sources, we estimated 3$\sigma$ upper limits of $J=19.2$ and $K_s=19.6$ (both Vega, and calibrated relative to 2MASS) for WIRC and FourStar, respectively. We compared the $J$ and the $K_s$ limits with empirical data as a function of spectral type for main sequence stars from \citep[updated on 2022~April~16]{2013ApJS..208....9P}, together with the three-dimensional extinction model from \citep{2019ApJ...887...93G}.  We found that for a nominal distance of 4.0\,kpc, we can exclude main sequence companion stars earlier than spectral type of about M3. 

Magnetars, a subclass of neutron stars with powerful magnetic fields and candidate progenitors for long-period radio transients \citep{GLEAM-X1627,BeniaminiPopulation}, are known for emitting periodic X-ray bursts \citep{MagnetarReview}. Therefore, we conducted observations to search for X-ray activity in \ulpo{}. Two observations lasting 0.8\,ks and 1\,ks were made with the Swift X-ray satellite \citep{xrt} on 2024-02-29 and 2024-06-02 during ASKAP and ATCA radio observations, respectively. Despite detecting radio pulses in both observations, we did not detect any X-ray bursts in the light curve binned at 2.5 \,seconds. An additional observation of 0.8 ks was conducted on 2024-03-01. We detected three raw photon counts from the stacked image using all three observations, and determined the upper limit of the X-ray luminosity to be $L_X \approx 7.4 \times 10^{32}\, \luminu$ (see Methods). This is comparable to the X-ray luminosities of magnetars ($\sim 10^{30} - 10^{35} \luminu$) \citep{MagnetarCatalogue} and similar to the X-ray luminosity upper limits of other known long-period radio transients ($\lesssim 10^{30} - 10^{32} \luminu$) \citep{GLEAM-X1627, GPM1839,ASKAPJ1935}. Additional X-ray observations were carried out with the Neutron star Interior Composition Explorer (NICER) \citep{NICER} instrument on-board the International Space Station. In total, 8 observations were conducted which lasted 0.2--4\,ks from 2024-07-12 to 2024-07-20 and resulted in a total exposure time of 13.9\,ks. Some observations were conducted simultaneously with ATCA or ASKAP, which had detected radio pulses. No X-ray emission was found in any observations, which gave us an upper limit on the peak X-ray luminosity of $\approx $3.0$\times$10$^{33}$ \luminu (see Methods).

The high brightness temperature of \ulpo{}, along with the microstructure and strong polarisation fractions, indicate that the radio emission mechanism is coherent and originates from a compact object. Consequently, we consider neutron stars and magnetic white dwarfs as leading progenitor candidates for \ulpo{} and discuss three possible scenarios below. 

\textit{\underline{Scenario 1}}: We first consider the scenario where the source is a rotationally-powered isolated magnetic white dwarf. We note that all known \emph{isolated} magnetic white dwarfs fall below the threshold required for coherent radio emission \citep{ReaPopulation}. Furthermore, the minimum magnetic field strength needed to generate the observed radio emission can be found by assuming curvature pair production at the polar caps, $B \approx 5.2 \times 10^{12} R^{-2} P^{3/2} \approx 5.2 \times 10^{11} \rm G$ \citep{PulsarDeathLine,ReaPopulation}, where $R = 6000$\,km is the radius of the white dwarf. This field strength is at least a factor of 100 stronger than the highest field ever detected in a white dwarf ($B \approx10^9 \rm G$ \citep{MHD_population}). Under the assumption that coherent radio emission originated from pair production, we can relate the source's radius to its period, magnetic field strength, and compactness \citep{BeniaminiPopulation,ASKAPJ1935} (see Methods). Even if we assume that \ulpo{} is a magnetically-powered white dwarf with a field strength close to the higher end of the observed distribution (i.e. $\approx10^9 \rm G$), the lower limit on the radius will be $0.6 R_{\odot}$, which is too large for a white dwarf.
While much dimmer radio emission is common amongst binaries containing a white dwarf and is interpreted as arising from the lower corona of the donor star \citep{Barrett2020}, radio emission has never been detected in isolated magnetic white dwarfs \citep{Pelisoli2024}. We finally note that whereas quasi-periodic oscillations with periods of 1-3\,s have been often observed in the optical flux of accreting magnetic white dwarfs \citep{BonnetBidaud2015QPOs}, such quasi-periodicity has never been detected in isolated objects. The above considerations further strengthen the view that an isolated magnetic white dwarf is highly unlikely to be the central engine of \ulpo{}.

\textit{\underline{Scenario 2}}: On the contrary, if \ulpo{} is a rotationally powered isolated neutron star and spins down purely through dipole radiation, the current spin down limit cannot appreciably constrain the B-field, as it will be as high as $B \sim 3.2 \times 10^{19} \sqrt{P\dot{P}} \lesssim 10^{18} \rm G$. This field strength is comparable to the maximum permissible poloidal field in a neutron star of $10^{18}$\,G \cite{Bocquet1995} and 1000 times higher than the highest field ever inferred in neutron stars (${\sim}10^{8}-10^{15}\, \rm G$ \citep{ATNFCatalogue}). 
Extended Data Figure \ref{fig:P_Pdot_deathline} shows the period derivative as a function of the spin period for different types of neutron stars and long-period radio transients. \ulpo{}, along with other long-period radio transients, fall below the theoretical death line for a pure magnetic dipole, where no radio emission is expected. Furthermore, the radio luminosity of \ulpo{} is an order greater than its spin-down luminosity. Given that the radio conversion efficiency from spin-down is less than 1\% \citep{RadioEfficiency}, it is clear that steady electromagnetic spin-down torques alone cannot explain the radio emission observed in long-period (minutes to hours) neutron stars. 
Another characteristic of \ulpo{} that strengthens its classification as a neutron star is its quasi-periodic substructure. Radio-loud magnetars, rotating radio transients, radio pulsars, and GLEAM-X\,J162759.5$-$523504.3 have been seen to exhibit quasi-periodic substructure in their emission \citep{PulsarQuasi-periodicity}. Ultimately, it is unclear what causes the quasi-periodicity in these objects but could be crucial to understanding the radio emission properties of neutron stars across the population. 


\textit{\underline{Scenario 3}}: Alternative possibilities include (1) orbital power in a tight binary where the compact object has a 6.45 hour spin period (possibly nearly-synchronized with the orbit \cite{1983ApJ...274L..71L}), or (2) a magnetically-powered slow neutron star that may or may not be in a binary \citep{2008ApJ...681..530P}. For (1), in the case of either a white dwarf or neutron star compact object with a 6.45 hour spin period, the unseen, low-mass companion must be well within the light cylinder of the compact object to provide sufficient magnetic torques on the orbit to power the observed emission. It is not immediately clear why the source flux would be dimming without a sizable $\lvert\dot{P}\rvert$ in a binary scenario without invoking additional ingredients. Moreover, the compact object must be exceptionally magnetised or the orbital period much shorter than 6.45 hours for binary interactions to power the observed emission. Thus, we disfavor this possibility. A more natural possibility, given the source’s variability characteristics, is an ultra-long-period magnetar, powered magnetically by either plastic motion on the crust or thermoelectric effects due to late time internal core field evolution \citep{BeyondDeathline-Alex&Zorawar}. Given the source luminosity, plastic motion greater than $10^2$ cm\,yr$^{-1}$ on a polar patch of the magnetar is sufficient to explain the observed characteristics. The timescale of dimming is also compatible with time scale of field twist evolution expected of plastic flows. This possibility may not generally lead to the interpulse and both poles being active, while thermoelectric effects from core evolution may be more symmetric with regard to hemispheres where pair production could occur. Finally, we note that the interpulse of \ulpo{} has implications for formation, as well as secular evolution of torques which may permit such an oblique rotator \citep{2018MNRAS.481.4169L}.

In summary, we report the discovery of the long-period radio transient \ulpo{}, which is the first of its kind to emit interpulses implying that the object is an oblique or orthogonal rotator with a rotation period of 6.45\,hour. Interpulses have been seen in radio pulsars and have played an important role in studying the emission beam structure \citep{Johnston_Props_of_IP, PSRB1702-19} and geometric parameters \citep{Manchester1977, Interpulse_AxisAlignment} of neutron stars. The polarisation features and evidence of radius-to-frequency mapping seen in \ulpo{} can provide more insights into the emission mechanism and viewing geometries of orthogonal rotators. We have shown that an isolated magnetic white dwarf or rotationally powered neutron star is unlikely to be responsible for the radio emission of \ulpo{}. The decaying radio luminosity and microstructure superimposed on the pulses strengthen our argument that the source is a magnetically powered neutron star either isolated or in a binary system. \ulpo{} share several properties with other known long-period radio transients, suggesting a common origin and emission mechanism across this class of objects. 
Although \ulpo{} shows a decreasing trend in radio luminosity, future monitoring may witness a rebrightening similar to GLEAM-X J162759.5$-$523504.3202. Population synthesis studies estimate that hundreds of long-period radio transients exist in our Galaxy, but they have been overlooked because traditional pulsar search methods are biased against long-period objects and wide pulses \citep{BeniaminiPopulation}. The discovery of \ulpo{} can improve our understanding of the population density of long-period radio transients in future studies, and the presence and properties of interpulse emission can test theoretical models of compact objects.

\begin{landscape}
\begin{table*}
    \centering
    \begin{tabular}{@{\extracolsep{\fill}} cccrrcrr }
    \toprule
    SBID & Start time & Duration & Central & Bandwidth & Telescope & \multicolumn {2}{c}{Peak flux density}\\
    & & & Frequency & & & MP & IP \\
     & UT & UT & MHz & MHz & & \multicolumn {2}{c}{Jy} \\
    \hline
     57929 & 2024-01-26 03:12:35 & 00:14:56 & 887.5 & 288 & ASKAP & 0.70 & - \\
     58387 & 2024-02-01 02:46:53 & 02:03:32 & 943.5 & 288 & ASKAP & - & 0.063\\
     58609 & 2024-02-03 21:03:04 & 08:05:35 & 943.5 & 288 & ASKAP & 1.4 & 0.063 \\
     58753 & 2024-02-06 20:51:42 & 08:09:31 & 943.5 & 288 & ASKAP & 1.0 & 0.13 \\
     20240212-0019 & 2024-02-18 04:00:57 & 08:01:45 & 816 & 544 & MeerKAT & 1.8 & 0.12 \\
     20240215-0217 & 2024-02-18 04:01:13 & 08:01:23 & 2187.5 & 875 & MeerKAT & 0.10 & 0.026 \\
     59605 & 2024-02-29 02:47:43 & 01:03:05 & 943.5 & 288 & ASKAP & - & 0.095 \\
     60091 & 2024-03-15 18:18:01 & 08:31:49 & 943.5 & 288 & ASKAP & 0.78, 0.78 & 0.37 \\
      - & 2024-04-29 14:05:15 & 07:49:50 & 2100 & 2048 & ATCA & 0.093 & 0.019 \\
      - & 2024-06-02 13:35:15 & 01:49:20 & 2100 & 2048 & ATCA & 0.023 & - \\
      - & 2024-06-15 11:25:55 & 00:58:30 & 2100 & 2048 & ATCA & 0.030 & - \\
      63296 & 2024-06-26 11:47:12 & 02:59:56 & 943.5 & 288 & ASKAP & 0.043 & - \\
      64280 & 2024-07-30 09:33:25 & 09:00:08 & 943.5 & 288 & ASKAP & 0.022 & - \\
      64328 & 2024-08-02 09:31:02 & 08:59:58 & 943.5 & 288 & ASKAP & 0.13 & - \\
      64345 & 2024-08-03 09:02:00 & 09:00:05 & 943.5 & 288 & ASKAP & 0.084, 0.060 & - \\
     \hline
    \end{tabular}
    \caption{All of the radio observations of \ulpo{}, including the scheduling block ID (SBID), start time and duration of each observation in UT, central frequency and band width in MHz, and the telescope used. The last two columns listed the peak flux density of the pulse(s), where MP indicates a main pulse and IP indicates an interpulse. There is no SBID for ATCA observations. We note that no interpulse was detected in recent ASKAP observations, even though the observation duration covered the entire period of the source.}
    \label{tab:ObservationDetails}
\end{table*}
\end{landscape}

\newcolumntype{P}[1]{>{\centering\arraybackslash}m{#1}}
\newcolumntype{M}[1]{>{\centering\arraybackslash}m{#1}}
\clearpage
\begin{table*}
    \centering
    \begin{tabular*}{1\linewidth}{@{\extracolsep{\fill}} P{6.3cm} M{4cm}}
         \hline \hline
         Parameter & Value \\
         \hline \hline
         \multicolumn{2}{c}{Measured properties} \\
         \hline
         Right ascension, $\alpha$ (J2000) & $18^h39^m50.573^s \pm 0.99''$ \\
         Declination, $\delta$ (J2000) & $-07\degree 56'39.170'' \pm 0.45''$ \\
         Period, $P$ & $23221.740 \pm 0.332$ seconds \\
         3$\sigma$ constraint on period derivative, $\dot{P}$ &  $<1.6\times 10^{-7}$ \pdotunits \\
         Observation epoch (in MJD) & 60335 -- 60525 \\
         Interpulse longitude separation from main pulse & $177.8\degree \pm 3.0 \degree$ \\
         Dispersion Measure, DM & $188.4 \pm 5.3 \, \dmunits$ \\ 
         Rotation Measure, RM & 214 -- 219 \rmunits \\
         Main pulse radio luminosity, $L_{\nu}$ &  $2 \times 10^{28} \luminu$\\
         Pulse width at half maximum at 1 GHz, $W_{50}$ & 320 - 710 seconds \\
         Main pulse linear polarisation fraction, $L/I$ & 60\% - 90\% \\
         Main pulse circular polarisation fraction, $V/I$ & 30\% - 60\% \\
         Interpulse linear polarisation fraction, $L/I$ & $\sim 90 \%$ \\
         Interpulse circular polarisation fraction, $V/I$ & $\lesssim 10 \%$ \\
         Inband spectral index, $\alpha$ & ${-}2.27\leq \alpha \leq {-}$2.99 \\
         \hline
         \multicolumn{2}{c}{Inferred properties} \\
         \hline
         Spin-down luminosity (neutron star), $\dot{E}$ & $\lesssim 10^{26} \luminu$ \\
         Spin-down luminosity (white dwarf), $\dot{E}$ & $\lesssim 10^{31} \luminu$ \\
         Surface dipole magnetic field strength (neutron star) & $\lesssim 3 \times 10^{18} \rm G$ \\
         Surface dipole magnetic field strength (white dwarf) & $\lesssim 5 \times 10^{11} \rm G$ \\
         X-ray luminosity upper limit, $L_X$ & $\lesssim 10^{33} \, \luminu$ \\
         Distance (YMW16) & 3.7 kpc\\
         Distance (NE2001) & 4.2 kpc \\
         \hline
    \end{tabular*}
    \caption{Key physical parameters measured or derived from observations. The errors represent the 3-$\sigma$ confidence interval.}
    \label{tab:sourceparams}
\end{table*}

\setcounter{figure}{0}
\captionsetup[figure]{name={\bf Extended Data Figure}}
\begin{figure}
    \centering
        \includegraphics[width=0.9\textwidth]{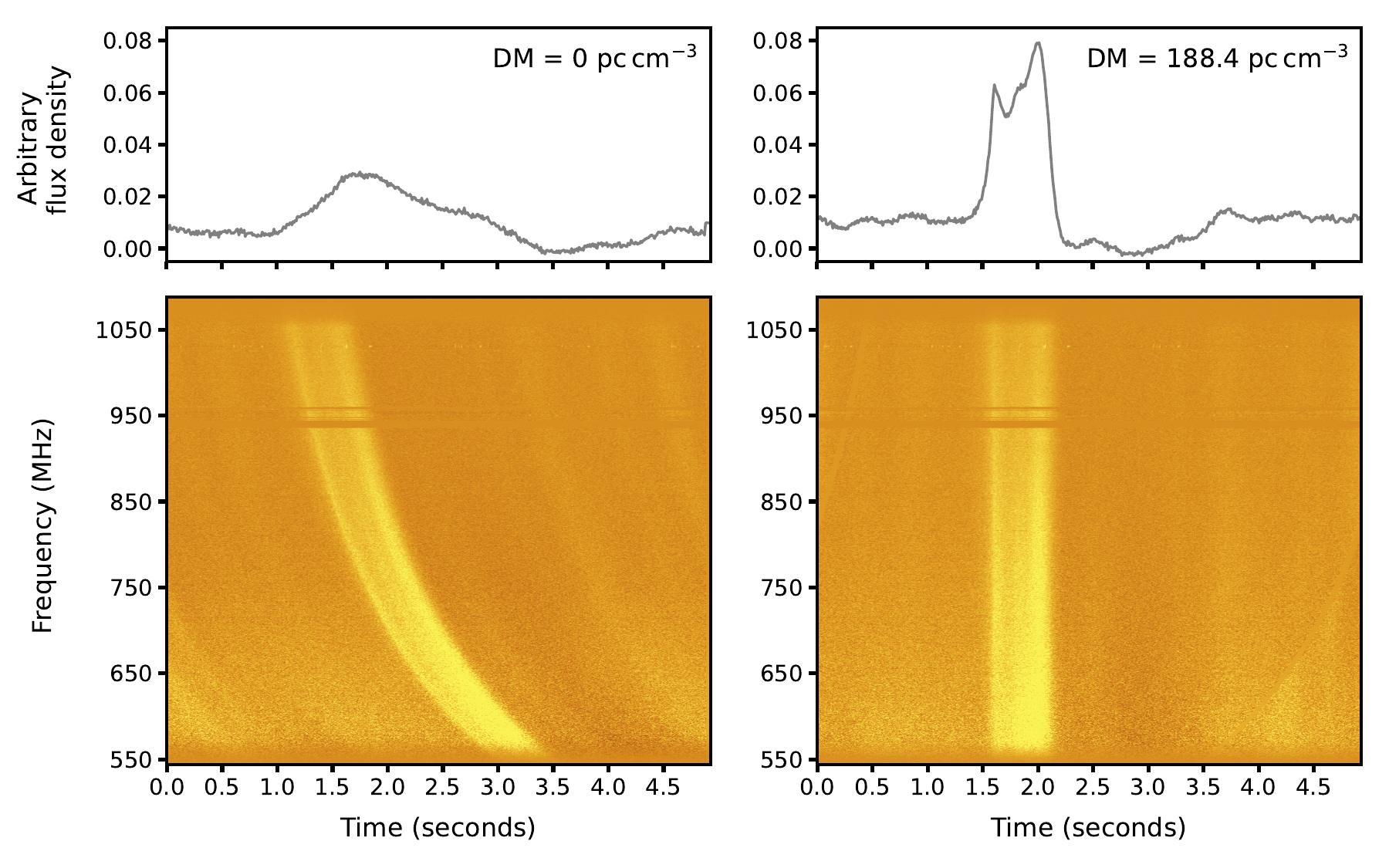}
    \caption{Light curves and dynamic spectrum of a sub-pulse detected in the MeerKAT UHF-band PTUSE data. The sub-pulse was observed on 2024-02-18 at 05:53:09 UTC, lasting approximately 500 ms. The left panels show the sub-pulse before de-dispersion, while the right panels show it after de-dispersion at a DM of 188.4 \dmunits.
}
\label{fig:subpulse}
\end{figure}

\begin{figure}
    \centering
    \includegraphics[width=\textwidth]{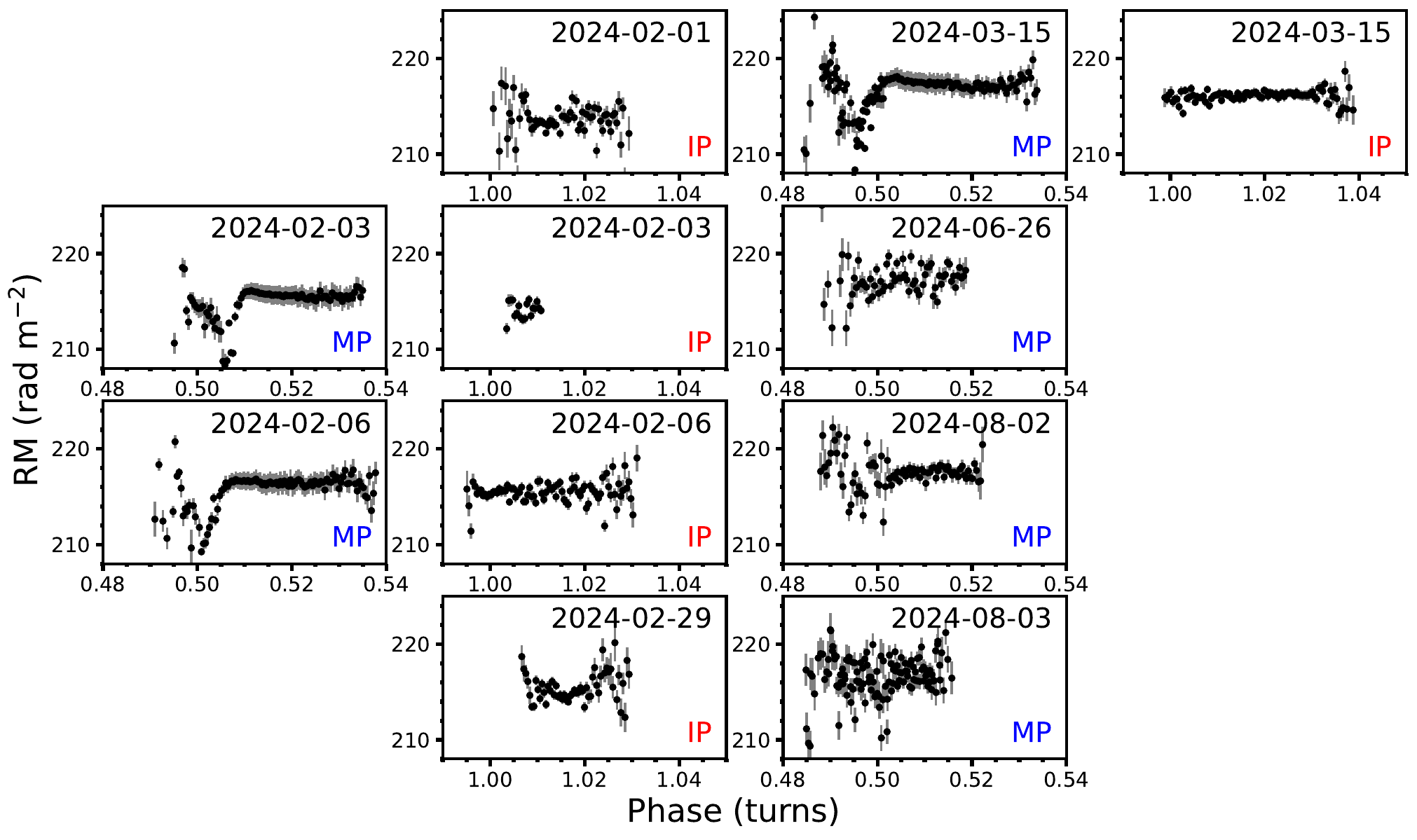}
    \caption{RM of the main pulses (MP) and inter-pulses (IP) of \ulpo{} as a function of pulse phase. Columns 1 and 3 show the main pulse, while columns 2 and 4 show the interpulse. Rows are arranged in order of date, with adjacent subplots in rows 1-2 and 3-4 corresponding to the same observing epoch. Missing axes are where there was no coverage or detection of the relevant pulse component.}
    \label{fig:rmvar}
\end{figure}

\begin{figure}
    \centering
    \includegraphics[width=0.825\textwidth]{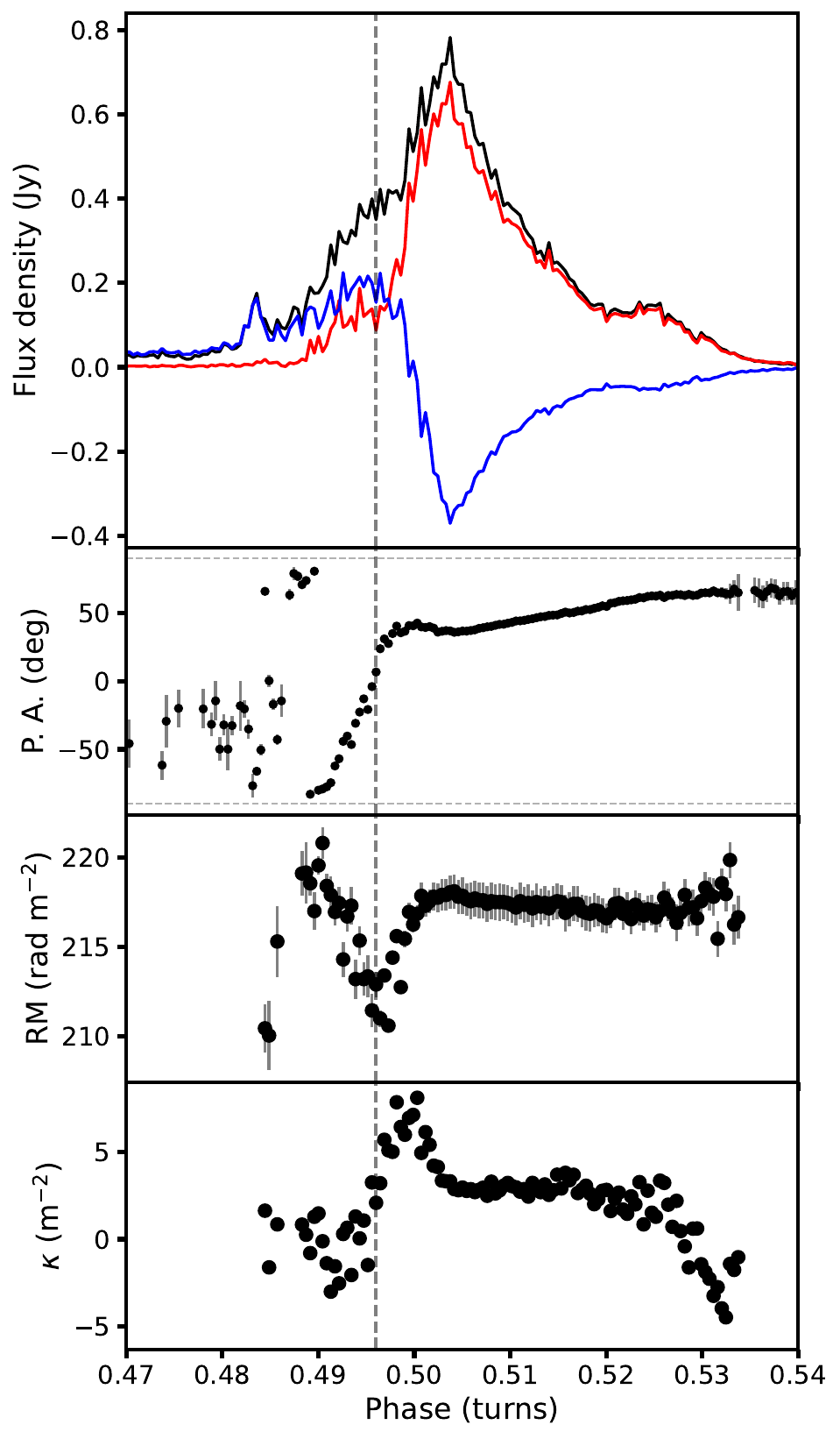}
    \caption{Variability in polarization properties of an example main pulse on 2024-03-15. The first and second rows show the total and polarized intensity, and the PA of the pulse, as in Figure \ref{fig:pulseprofile}. The third row shows the RM, and the fourth row shows the first-order frequency-dependence of $V/I$ over $\lambda^2$, $\kappa = \partial (V/I) / \partial(\lambda^2)$. To guide the eye, we have placed a vertical dashed line at the phase corresponding to the steepest variability in $\kappa$.}
    \label{fig:rm_v_var}
\end{figure}

\begin{figure}
    \centering
    \includegraphics[width=\textwidth]{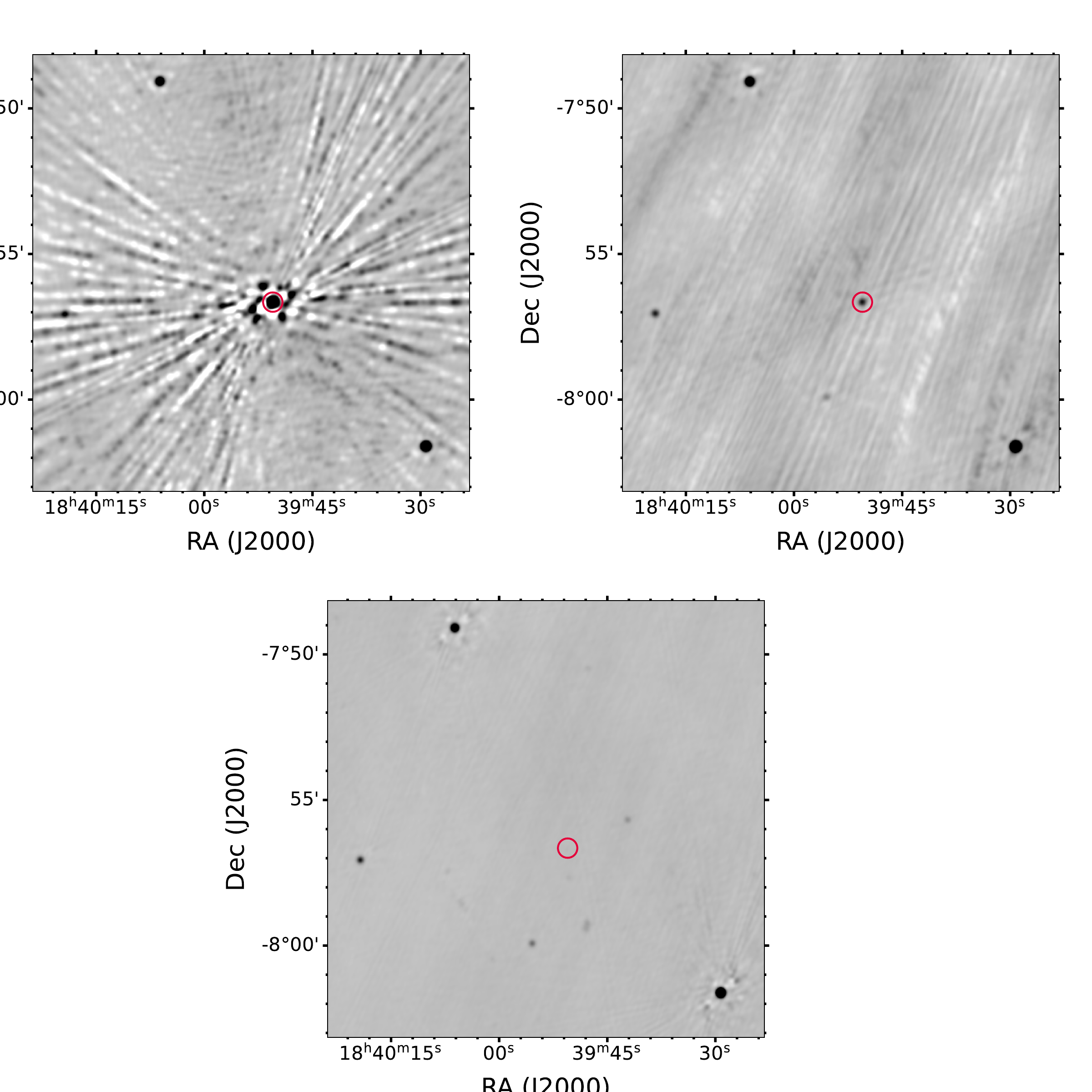}
    \caption{The MeerKAT images at the coordinates of \ulpo{}. The upper left image shows a main pulse, while the upper right shows an interpulse. Both images are snapshots lasting for 30 minutes around the pulse ToA. The bottom image is a 2.5-hour deep image between the main pulse and interpulse (i.e., off-pulse period) with an RMS of ${\sim}90 \, \mu \rm Jy/beam$. The red circle in each image indicates the position of the source and does not carry any physical meaning on the accuracy of the position.}
    \label{fig:MeerKAT_Images}
\end{figure}

\begin{figure}
    \centering
        \includegraphics[width=0.75\textwidth]{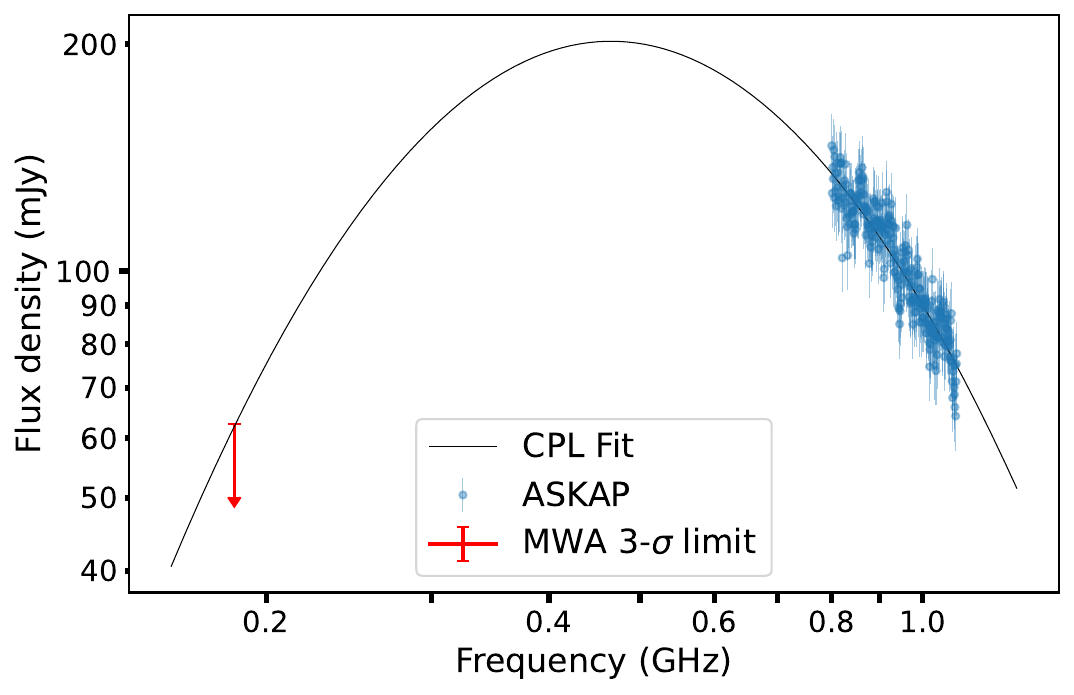}
        \captionsetup{singlelinecheck=off}
    \caption[.]{Simultaneous broadband measurements of the pulse at 2024-08-02 14:30 observed with the MWA and ASKAP. All data have been corrected for the primary beam. A curved power-law fit of the form 
    $S_\nu \propto \nu^\alpha \exp{[q \left(\log{\nu}\right)^2]}$ has been fit to the data, finding $\alpha=-2.11\pm0.07$, $q=-1.38\pm0.06$, and $S_\mathrm{1GHz}=90.1\pm0.6$\,mJy. Since the MWA point is only a limit, the fitted curvature $q$ value is an upper limit.
    \label{fig:radio_SED}}
\end{figure}

\begin{figure}
    \centering
    \includegraphics[width=1\linewidth]{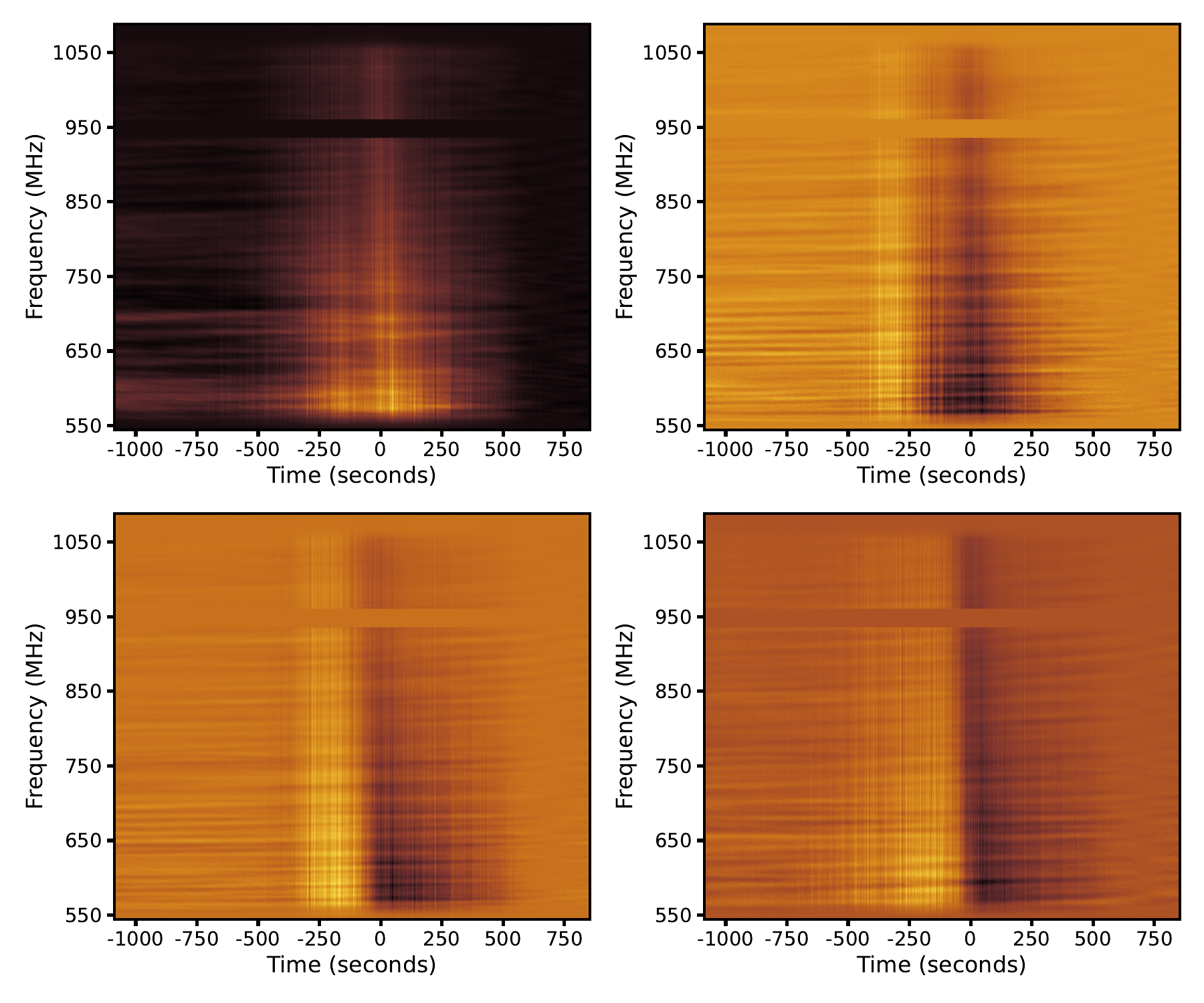}
    \caption{Dynamic spectrum of the MeerKAT UHF-band backend data with a time resolution of 60.24 $\mu \rm s$ for Stokes I (top left), Q (top right), U (bottom left), and V (bottom right) parameters. The horizontal bright stripes at 600 MHz and 700 MHz are radio frequency interference and baseline variation that we were unable to mitigate. These stripes were also seen during off-pulse period. The zebra-like horizontal patterns are likely to be baseline variations that were also seen at different frequency range during off-pulse period. We therefore conclude that these patterns are not intrinsic to the source.}
    \label{fig:MeerKAT_Stokes_Uband}
\end{figure}

\begin{figure}
    \centering
    \includegraphics[width=1\linewidth]{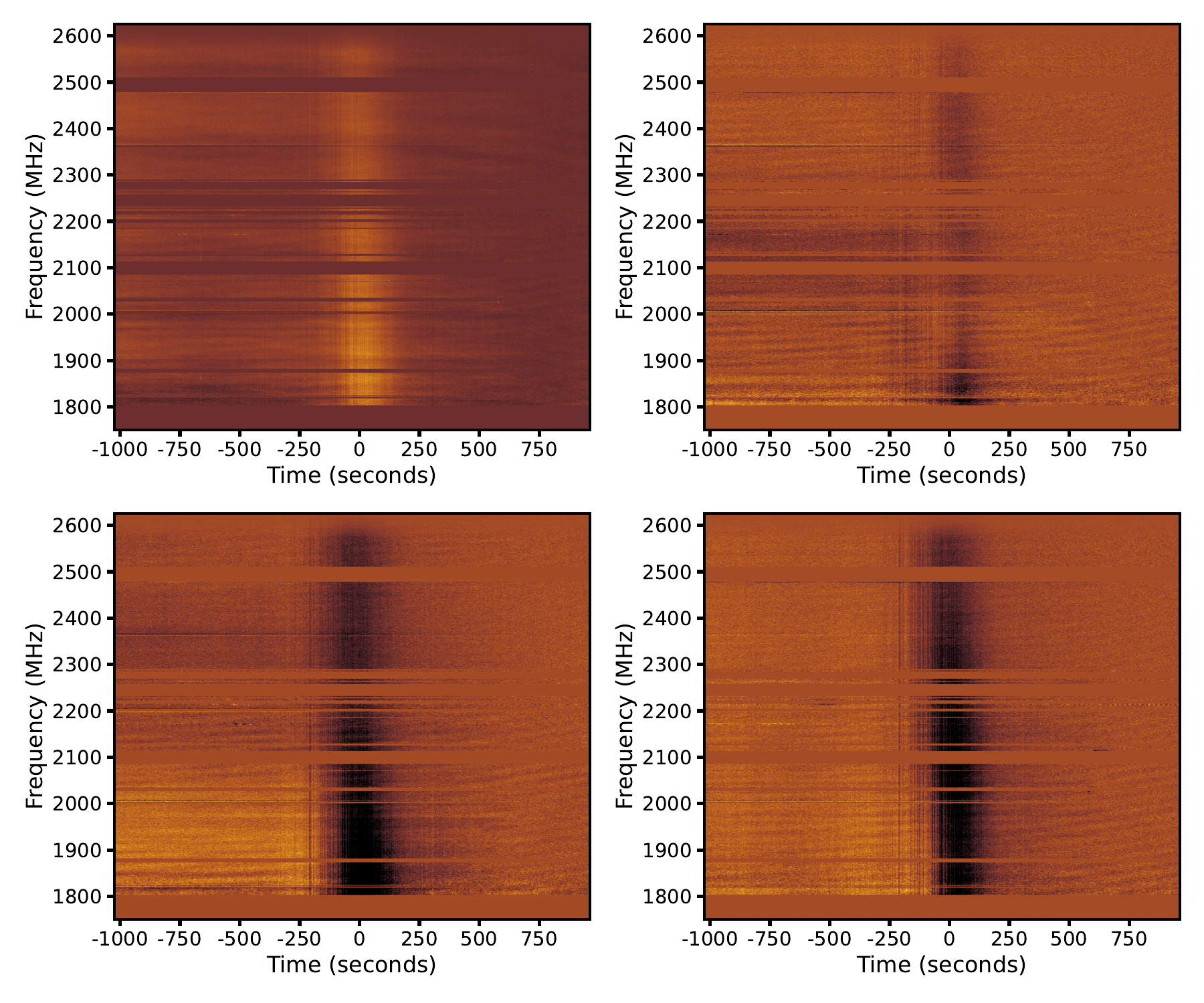}
    \caption{Dynamic spectrum of the MeerKAT S-band backend data with a time resolution of 37.45 $\mu \rm s$ for Stokes I (top left), Q (top right), U (bottom left), and V (bottom right)  parameters. Horizontal strips similar to the previous figure are also seen in the dynamic spectrum and we determine that they are not intrinsic to the source.}
    \label{fig:MeerKAT_Stokes_Sband}
\end{figure}

\begin{figure}
    \centering
        \includegraphics[width=0.9\textwidth]{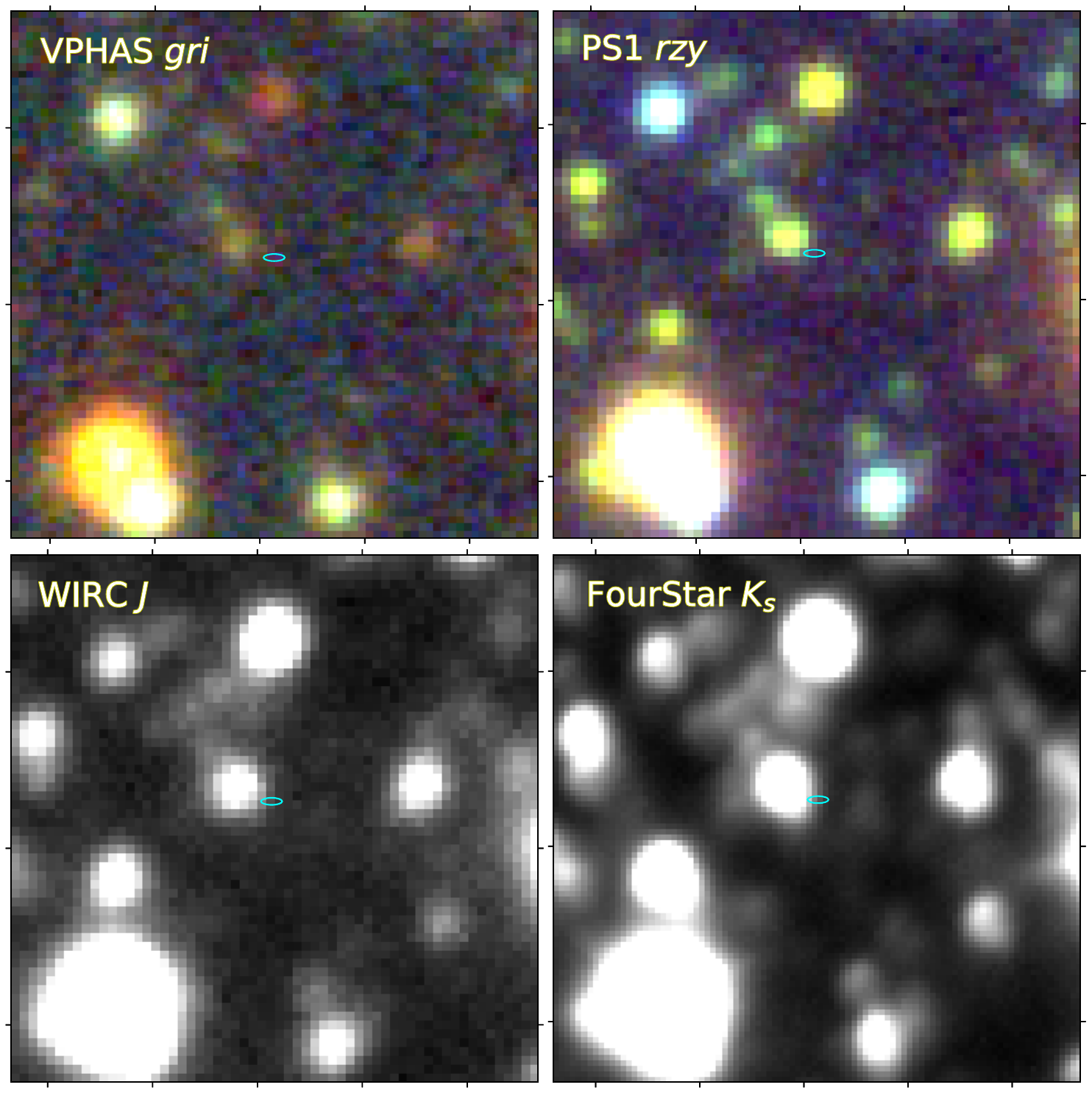}
    \caption{Optical and near-infrared images of the field of \ulpo{}. Each panel is centered at the source and $15^{\prime \prime}$ on a side with north up and east to the left.  We show the same VPHAS $gri$ composite, a Pan-STARRS $rzy$ composite, and our WIRC $J$-band and FourStar $K_s$-band images from left to right.  In those images we show the best-fit radio position as the cyan error ellipse.}
    \label{fig:OIR}
\end{figure}

\begin{figure}
    \centering
        \includegraphics[width=0.9\textwidth]{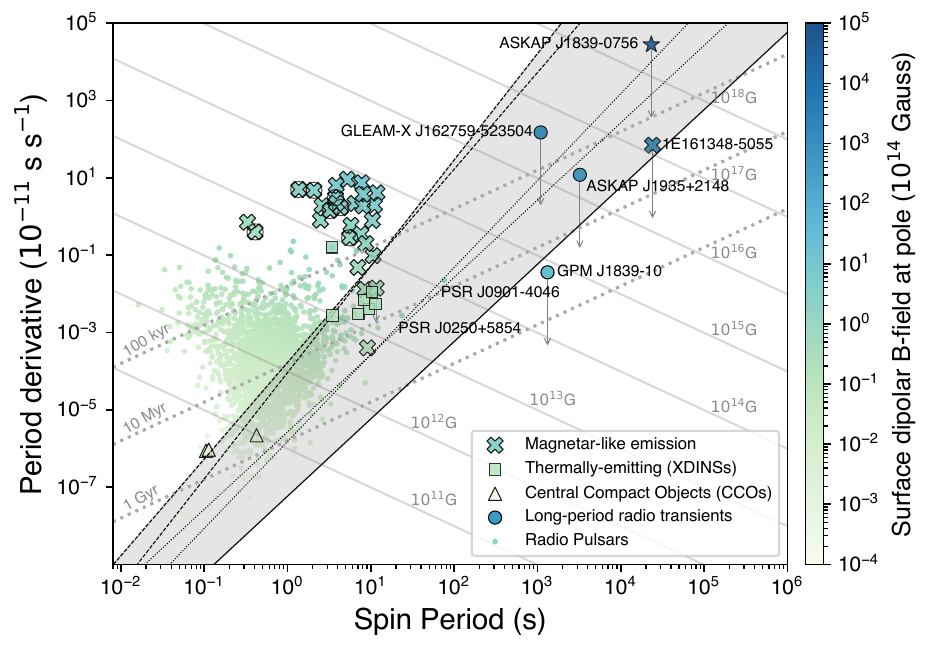}
    \caption{A period-period derivative diagram showing the spin-period against the period derivative for different types of neutron stars and compact objects. The plot is adapted from \cite{GPM1839}. The points are colour-coded according to their surface dipolar magnetic field ($B = 3.2 \times 10^{19} \sqrt{P \dot{P}}$ G). The dashed lines indicate the theoretical death lines for a pure dipole, dotted lines for a twisted dipole, and solid lines for the twisted multipole configuration \citep{PulsarDeathLine, Zhang+2000_deathline}. Below these lines, no radio emission is expected.
}
    \label{fig:P_Pdot_deathline}
\end{figure}


\begin{figure}
    \centering
        \includegraphics[width=0.75\textwidth]{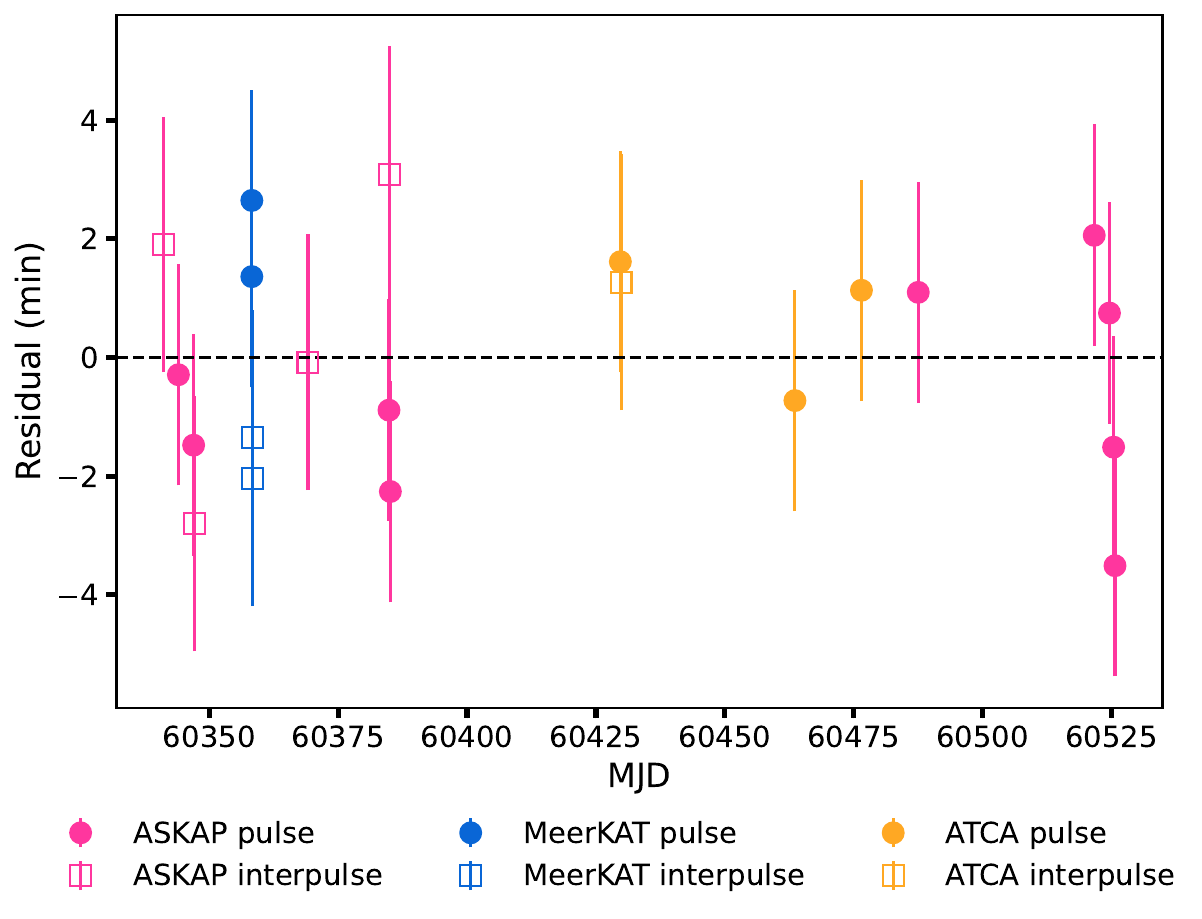}
    \caption{Timing residuals for \ulpo{}.  Data are shown for the main pulse (filled circles) and interpulse (open squares), where the later have been shifted by the best-fit interpulse interval of $11466$\,s (or $177.8^\circ$) relative to the main pulse.  Data from ASKAP are shown in pink, data from MeerKAT are shown in blue, and data from ATCA are shown in orange.}
    \label{fig:resids}
\end{figure}

\clearpage

\section*{Methods}\label{sec6}

\section*{ASKAP}
\label{sec:askap}

The ASKAP array comprises 36 antennas each of which is equipped with a prime-focus phased array feed (PAF). This allows the array to form 36 beams on the sky that cover 30 $\rm deg^2$. The signal is channelised to 1 MHz and has a bandwidth of 288 MHz. The minimum integration time of ASKAP is 10 seconds. \ulpo{} was first found in a routine RACS low-band pointing \citep{RacsLow-McConnell,RacsLow-Duchesne} centered at 943.5 MHz. The source appeared as a peak in the Stokes V (circular polarization) image without any known counterparts. Subsequent inspection of the dynamic spectrum revealed the source to be strongly linear polarised. Sub-pulses within the burst were discovered independently by the CRAFT Coherent (CRACO) backend \citep{wang2024craftcoherentcracoupgrade} which were studied in array-coherent filterbank data simultaneously at a 13.8~ms time resolution.

On-axis leakage calibration is performed by the processing pipeline using the \texttt{cbpcalibrator tool} in \texttt{ASKAPSOFT}. This determines the leakages after finding the bandpass solution from observations of the unresolved and unpolarised primary flux calibrator PKS B1934-638 \citep{reynolds1994revised}. The derived solutions (e.g. \citep{1996A&AS..117..137H,1996A&AS..117..149S}) are applied directly to the bandpass-calibrated and self-calibrated visibilities. This procedure is basically identical to other ASKAP scientific project observations, such as the ASKAP Variables and Slow Transients and Evolutionary Map of the Universe Survey. The residual on-axis instrumental leakage level is typically less than 0.1\%.

Ten ASKAP Target-of-Opportunity (ToO) observations conducted between 2024-02-01 and 2024-08-03 were performed in the \texttt{closepack36} configuration, with 0.9 deg pitch and a 943.5 MHz central frequency. For observations up to 2024-06-26, the source was positioned at the center of beam 27, allowing coverage of nearby sources of interest in other beams. In the remaining three observations, the telescopes were centered at other objects and \ulpo{} was placed in beam 34 (SBID 64280), beam 27 (SB64328), and beam 28 (SB64345). All data were processed using the \texttt{ASKAPSOFT} pipeline to subtract background sources. A light curve and dynamic spectrum at the source's coordinates were generated for each epoch to search for pulses. The rotation measure (RM) of the pulses was determined by brute-force searching for a peak signal-to-noise ratio in linear polarization, and the polarization position angle (PA) was calculated based on the de-Faraday rotated Stokes Q and U parameters. 

\section*{ATCA}
ATCA consists of 6 antennas capable of observing in bands with frequencies ranging from 1.1 GHz to 105 GHz. All ATCA observations in this study were conducted in the L-band, spanning frequencies from 1.1 GHz to 3.1 GHz. The array configurations on 2024-04-29, 2024-06-02, 2024-06-15 were 6A, H168, and 6D, respectively. The array configuration does not impact the observations, as our goal was to extract the pulse arrival time from the light curve rather than to image the source. All ATCA observations used J1939\ensuremath{-}6342 as the flux calibrator and primary calibration source 1819\ensuremath{-}096 as the phase calibrator. The data were reduced using \texttt{Miriad} \citep{Miriad}, and \texttt{CASA} \citep{CASA} was used to generate the light curve and dynamic spectrum.

\section*{MeerKAT}

The Meer(more) Karoo Array Telescope (MeerKAT) operated by the
South African Radio Astronomy Observatory (SARAO) comprises 64, 13.5-m
antennas distributed over 8-km in the Karoo region in South Africa. 40 of these dishes are concentrated in the inner $\sim1$-km core.
The MeerKAT DDT observation was conducted on 2024 Feb 18, in dual frequency sub-array mode (UHF-band centered at 816\,MHz and S0-band centered at 2187.5\,MHz) with 2-second correlator dumps. We used J1939\ensuremath{-}6342 as the flux calibrator and band-pass calibrator and J1833\ensuremath{-}2103 as the gain calibrator. The Pulsar Timing User Supplied Equipment \citep[PTUSE;][]{ptuse} backend of the MeerKAT pulsar timing project was also used to record the full polarization data. The PTUSE data was  recorded in the psrfits format with a time resolution of 60.24 $\mu \rm s$ in UHF-band and 37.45 $\mu \rm s$ in S0-band.

We refer readers to \citep{2021MNRAS.505.4483S} for a detailed review on the array calibration of MeerKAT. Broadly speaking, the process involves a delay calibration, where calibration products are calculated after observing the flux calibrator, and a phase up, where phase and bandpass correction are performed. We obtained gain and cross-phase solutions that were applied to remove leakage to get a calibrated beam at the pointing centre.  The polarization purity characteristics of the MeerKAT L-band receiver, including ellipticity, non-orthogonality, and differential ellipticity, are close to ideal. The antenna-based leakage terms, which describe the discrepancy between the system's response and a polarized signal, is negligible. Overall, the calibration method used provides reliable and accurate polarization properties.

The 2-second data was calibrated and flagged using \texttt{oxkat} \citep{oxkat}, and imaged around the target at 30-minute interval using \texttt{WSclean} \citep{wsclean} to determine the approximate time of arrival (ToA) of the pulse. The PTUSE data around the ToA were then processed using \texttt{PSRCHIVE} \citep{Psrchieve} to obtain the detailed profile and polarimetry of the pulse. The dispersion measure (DM) of the source is estimated by de-dispersing a sub-pulse in the MeerKAT UHF-band PTUSE data. The sub-pulse was found at 2024-02-18 05:53:09 UTC, which is near the centre of the full pulse, and lasted for approximately 500\,ms. The sub-pulse was de-dispersed using the \texttt{pdmp} tool in \texttt{PSRCHIVE}, resulting the best-fitted DM of $188.4 \pm 1.7 \dmunits$ with 1-$\sigma$ uncertainty. Taking the mean distance inferred from the NE2001 and YMW16 electron density model, we determined the distance of \ulpo{} to be 4.0 kpc. \citep{2021PASA...38...38P} compared the model-estimated distance with the parallax distance and found that the uncertainties in both models are approximately 30\%. Therefore, we conclude that the error in the distance measurement is 1.2 kpc. We note that the distance estimated from the DM systematically underestimate the true distance to the source. \citep{2023MNRAS.525.3963K}.

\section*{Period and period derivative estimation}
We made use of the Stokes I dynamic spectra from the various observations, some of which relied on \texttt{DStools}~\footnote{\url{https://github.com/joshoewahp/dstools/tree/main/dstools}} for extraction.  For each observation we identified the initial pulse time(s) based on a local maximum.  Once that was identified, we used a window of $\pm P/2$ to select a region for detailed analysis.  We determined the baseline (which can vary due to e.g., unsubtracted continuum sources) by excluding a 10\,min window around the pulse and fitting a low-order polynomial to the remaining data, separately for pre- and post-pulse.   The baseline was then interpolated across the pulse and subtracted from the lightcurve.

With the baseline-subtracted lightcurve, we modeled the pulse profile using 1--4 Gaussian components, depending on the complexity of the emission identified by eye.  The noise level for the fit was determined by the off-pulse background rms.  We fit for the components using \texttt{scipy.curvefit}.  In the case of multiple components, the pulse arrival time was determined arbitrarily to be the unweighted mean of the different components. The arrival time, pulse width, and flux density along with associated measurement uncertainties were recorded for further analysis.

These arrival times served as the basis for our timing analysis.  We used \texttt{PINT}\cite{PINT} to convert the topocentric arrival times to the pulsar time-of-arrival (TOA) format.  We determined a timing ephemeris, where we fixed the position to the best-fit position from the lightcurve and used the initial separation between the two pulses in the 2024-03-15 ASKAP observation as the period.   We modeled the interpulse using a \texttt{JUMP} component, which establishes a constant time offset between the interpulse and main pulse.  The dispersion measure was taken from the ASKAP 13.8~ms time resolution filterbank data on 2024-01-26.  

The measurement uncertainties on the TOAs were typically $\approx 2\,$s, varying from $0.2$\,s to 6\,s.  However, these were far smaller than the observed scatter between TOAs assuming the ephemeris we derive.  Therefore we included an additional quadratic error term, \texttt{EQUAD} on each TOA, which we modeled separately for the main pulse and interpulse (we kept these values fixed between all telescopes, although changing that would not alter our results). Based on the observed scatter, we took initial values of ${\rm EQUAD}_{\rm MP}=50\,$s and ${\rm EQUAD}_{\rm IP}=100\,$s.  Again, these values are subject to revision but give us an initial estimate for the timing behavior. EQUAD is a parameter used in pulsar-timing that serves as a white noise or jitter noise \citep{10.1093/mnras/stw179,Reardon_2023}. From figure \ref{fig:light curve}, we see that the source showed multiple peaks in some pulses, leading to poor estimates on the ToA and underestimation of the fitting uncertainty. Therefore, this parameter acts as an empirical parameter to account for the variation in the pulse shape and width. Given that EQUAD ({$\sim$}100s) is considerably smaller than the pulse width (${>}$300s)and 0.5\% of the period, we believe that the correction is reasonable.

We then fit the ephemeris in \texttt{PINT}.  The free parameters were the frequency $F_0$ and the interpulse jump.  We obtained a reasonable fit, with $\chi^2=70.4$ for 18 degrees-of-freedom.  To refine our assumed uncertainties (modeled as \texttt{EQUAD}), we refit the MP and IP data separately, and scaled the \texttt{EQUAD} values for each by the square-root of the reduced $\chi^2$ to ensure a final reduced $\chi^2$ near one.  This resulted in ${\rm EQUAD}_{\rm MP}=111.8\,$s and ${\rm EQUAD}_{\rm IP}=129.2\,$s, not too far from our initial estimates and similar to each other, suggesting comparable levels of timing variation.  With these values our final fit had $\chi^2=18.3$ for 18 degrees-of-freedom.  See Figure~\ref{fig:resids} for the timing residuals.

There is no sign of any secular variation in the residuals, nor any other obvious pattern, but the small number of measurements together with pulse-to-pulse variability makes this unconstraining.  If we allow for a first derivative of spin frequency, the $\chi^2$ only changes by a small amount,  $-1.2$, indicating that this there is no statistically significant spin-down or spin-up.  From this we determine a frequency derivative $F_1=(1.1\pm 1.0) \times 10^{-16}\,{\rm Hz\,s}^{-1}$.  This gives a $3\sigma$ constraint on the period derivative of $\lvert\dot{P}\rvert<1.6\times 10^{-7}\,{\rm s\,s}^{-1}$. From this we can infer limits to the spin-down luminosity, characteristic age, and magnetic field as given in Table~\ref{tab:sourceparams}. 

\section*{Radio luminosity and Inband spectral index}
The spectral index is estimated by fitting a power law to the flux intensity of the main pulse against the frequency. For ASKAP observations where the source was not centered in the beam, a primary beam-corrected spectral energy distribution (SED) is used for fitting after correcting the measured SED using a Gaussian function. The spectral index from ASKAP observations varies between $-2.27 \pm 0.06$ and $-2.99 \pm 0.11$. However, MeerKAT and ATCA observations are subject to greater baseline variations and radio frequency interference, which compromised the goodness of fit. We smoothed those data by rebinning along the frequency axis and found the spectral index to be $-2.65 \pm 0.02$ in the MeerKAT UHF-band and $-2.39 \pm 0.05$ in the S-band. The ATCA observations yielded a spectral index ranging from $-2.36 \pm 0.04$ to $-2.58 \pm 0.03$ for the main pulses. The spectral indices of each pulse were used to scale the flux density at 1400 MHz and plotted against time, as shown in Figure \ref{fig:decreasinglumin}.

By fitting a curved power-law same as in \citep{GPM1839}
\begin{equation}
    S_\nu = S_{\rm 1GHz} \left( \frac{\nu}{1 {\rm GHz}} \right)^\alpha \exp{ \left[ q \left(\log{\nu}\right)^2 \right] }
\end{equation}
to the ASKAP observation on 2024-08-02 and the upper limit of MWA observation, we found a turnover in spectral index at $\sim 300 - 600$~MHz (see Extended Data Figure \ref{fig:radio_SED}). The best-fit parameters are $\alpha=-2.11\pm0.07$, $q=-1.38\pm0.06$, and $S_\mathrm{1GHz}=90.1\pm0.6$\,mJy. Since the source lies at higher Galactic latitude than almost all known \textsc{Hii} regions \citep[and slightly beyond the coverage of the most comprehensive \textsc{Hii} region catalogue][]{2014ApJS..212....1A}, free-free absorption from such a region seems unlikely; however, a cold molecular cloud could produce such a turnover without being detected, as has been suggested as a possibility for some pulsars \citep{2016MNRAS.455..493R}. The turnover could also be intrinsic, potentially from synchrotron self-absorption in the magnetosphere \citep{1973A&A....28..237S}, or due to the nature of the the unknown emission mechanism. Simultaneous broadband observations covering the 200--700\,MHz would be necessary to fit physical models.

We calculated the radio luminosity of \ulpo{} by assuming a beam geometry similar to a radio pulsar,
\begin{equation}
    L = 4\pi D^2 \int {\rm sin}^2 \left[ \frac{\rho(\nu)}{2} \right] S_\nu {\rm d}\nu,
\end{equation}
where $D=4.0\, \rm kpc$ is the distance, $\rho$ is the beam opening angle and $S$ is the flux density \citep{Handbook_of_Pulsar,GPM1839}. Following the same treatment in \citep{GPM1839}, we assume that the beam opening angle can be approximated by a power law with index $\beta$. The luminosity can then be written as
\begin{equation}
    L = 4\pi D^2 \Omega_{1{\rm GHz}} \int \left( \frac{\nu}{1 \rm GHz} \right)^\beta S_\nu {\rm d}\nu,
\end{equation}
where $\Omega_{1{\rm GHz}} \approx \rho^2_{1 {\rm GHz}}/4$ is the solid angle at 1\,GHz and $\beta = -0.26$ is a typical value for radio pulsars \citep{Emission_Height}. Substituting equation (1) into the expression above and integrating over all frequencies gives us the following expression,
\begin{equation}
    L = 4\pi D^2 S_{\rm 1GHz} \Omega_{1{\rm GHz}} \sqrt{-\frac{\pi}{q}} {\rm exp} \left[ - \frac{(\alpha+\beta+1)^2}{4q} \right].
\end{equation}

The threshold, or death line, for radio emission from rotationally powered pulsar can be formulated by requiring the polar cap radius, $R_p$, to be greater than the curvature photon gap height, $\ell_{\rm gap}$ near the surface \citep{1979ApJ...231..854A,1981ApJ...248.1099A,BeniaminiPopulation}. The polar cap radius of our source can be calculated by 
\begin{equation}
    R_p \simeq \sqrt{\frac{2\pi R^3}{cP}} \sim 100 \, \rm cm,
    \label{eq:PC}
\end{equation}
where $R=10^6 \,$cm is the radius of the neutron star and $P$ is the period in second \citep{Handbook_of_Pulsar}. We adapted the formula of $\ell_{\rm gap}$ from \citep{gap_height} and omitted order unity factors,
\begin{equation}
    \ell_{\rm gap} \simeq 2 \times 10^4 \left(\frac{\rho_c}{10^7\,{\rm cm}} \right)^{2/7} \left(\frac{P}{1\,{\rm s}}\right)^{3/7} \left(\frac{B}{10^{12}\,{\rm G}}\right)^{-4/7}\, \rm cm \gtrsim 500\, cm,
\end{equation}
where $\rho_c= 10^7\,$cm is the radius of curvature and $B=10^{18}\,$G is the upper limit on the magnetic field. The gap height is greater than the polar cap radius, implying that the source is not solely rotation-powered. 

The polar cap radius and emission height, $r_{\rm em}$, is related by
\begin{equation}
    \frac{\sin^2\theta_{\rm em}}{r_{\rm em}} = \frac{\sin^2\theta_{p}}{R},
\end{equation}
where ($r_{\rm em}, \theta_{\rm em}$) is the coordinates of the emission point and $\sin \theta_{p} = R_p / R$. The emission height can be expressed as 
\begin{equation}
     r_{\rm em} \simeq 4 \times 10^7 \left ( \frac{\nu}{1 \rm GHz} \right)^{\beta} \left( \frac{\dot{P}}{10^{-15} \pdotunits} \right)^{0.07} \left( \frac{P}{\rm s} \right)^{0.3} \, {\rm cm} \sim 3 \times 10^9 \, {\rm cm},
     \label{eq:rem}
\end{equation}
where we take $\nu = $1\, GHz \citep{Handbook_of_Pulsar}. By requiring the true polar cap radius be larger than Eq.~(\ref{eq:PC}) and at least as wide as the gap height, we have $\theta_{\rm em} \gtrsim 0.03$. This implies an activated surface region much larger \citep[e.g.,][]{BeyondDeathline-Alex&Zorawar} than possible in a solely rotation-powered scenario. The beam opening angle can be approximated by $\rho \approx 3\theta_{\rm em}/2 \gtrsim 0.045$, which gives us the beam solid angle of $\gtrsim 5 \times 10^{-4}$. The opening angle can also be compared with the value predicted from the pulse width through spherical trigonometry,
\begin{equation}
    \cos \rho = \cos \phi \cos \gamma + \sin \phi \sin \gamma \cos \left( \frac{W}{2} \right),
\end{equation}
where $\phi$ is the angle between the rotational axis and the magnetic axis, $\gamma$ is the angle between the rotational axis and our line of sight, and $W$ is the pulse width in longitude of rotation \citep{Handbook_of_Pulsar}. If we assume $\phi \sim \gamma \sim \pi/2$ for an orthogonal rotator and a typical pulse width of 500\,s, the beam opening angle will be $\rho \approx 0.068$, which is in good agreement when using the pulsar scaling relation Eq.~(\ref{eq:rem}).

By substituting the best-fit parameters and considering that the main pulse flux density in February 2024 is an order brighter than that in August 2024, we report the beaming-corrected radio luminosity of the source to be $L \gtrsim 2 \times 10^{28} \,\luminu$.

\section*{Rotation Measure analysis}

Apparent intra-pulse variability in rotation measures is known to occur in some pulsars, magnetars, and fast radio bursts \cite{2015MNRAS.449.3223D, 2023PhRvD.108d3009K, ljl+23} and is attributed either to incoherent superposition of radiation from quasi-orthogonal modes, temporal smearing due to interstellar scattering, or generalized Faraday Rotation \cite{2004ApJ...606.1167R, 2009MNRAS.396.1559N,2019MNRAS.483.2778I}. 

Motivated by this, we measured the RM($\phi$) across each pulse detected in ASKAP observations using RM Synthesis \citep{bdb2005}, in the full 10\,s resolution ASKAP dynamic spectra. The results are shown in Extended Data Figure \ref{fig:rmvar}. For early epochs, we observe variability in the RM of about $\sim~5$\,rad\,m$^{-2}$ in the leading edge of the main pulse (MP) component. This variability can be broadly described as a dip and then rise in the RM leading into the trailing edge of the MP, which shows a more stable RM. On the other hand, the interpulse (IP) components exhibit relatively uniform RM over their duration. In later epochs (2024-06-26, 2024-08-02, and 2024-08-03), the dip and rise feature in the MP RM has disappeared. 

Propagation effects both in the source magnetosphere and along the ISM path are predicted to also induce changes in the apparent circular polarization fraction, $V/I$, across frequency \cite{2019MNRAS.483.2778I}. Motivated by the apparent variability in RM, we investigated the variability in the frequency dependence of $V/I$ across single pulses, measured by the first derivative of $V/I$ over $\lambda^2$, $\kappa = \partial (V/I) / \partial(\lambda^2)$, which we measure using a least-squares fit to a linear function for $V/I$ over $\lambda^2$. We show the results in Extended Data Figure \ref{fig:rm_v_var}, where variability of $\kappa$ correlated with RM variability is evident. This variability is strongest at pulse phase $\sim 0.496$, where the RM reaches a local minimum and $V/I$ is close to its maximum.

We also investigated the frequency dependence of the Stokes parameters using the MeerKAT data. The Stokes Q, U, V parameters are first normalized by the total polarized intensity and binned along the time axis at $\sim 15$ -second intervals. Each binned Stokes parameter was then plotted against the observing frequency, and an animation was created by stacking these plots in chronological order across the pulse. The animation was visually inspected to identify any notable frequency dependence in the polarization fractions. We observed a weak frequency dependence in the Stokes parameters. However, this dependence is not strong enough to rule out Faraday Conversion or wave mode coupling as the cause of such RM variability.

\section*{Model constraints}
\subsection*{White Dwarf}
The conventional model of coherent emission from compact object assumes that a vacuum gap exists above the polar cap. The potential difference across this gap has to be sufficiently large to sustain pair production. Following the discussion in \citep{BeniaminiPopulation} and the references therein, the death line for coherent radio emission can be formulated by relating the source's radius with the source curvature radius, rotational period, and magnetic field,
\begin{equation}
    R \gtrsim 4 \times 10^9 \left(\frac{Q_{c}}{10}\right)^{4/17}\,\left(\frac{P}{1000~\rm{s}}\right)^{13/17}\,\left(\frac{B}{10^9~\rm{G}}\right)^{-8/17}\,\rm{cm},
\end{equation}
where $P$ is the period, $B$ is the magnetic field and $Q_{c} = \rho_{c}/R$ is the dimensionless characteristic field curvature radius. For long-period radio transients, which has a smaller polar cap radius, $Q_{c} \gg 10$ is a more realistic choice. However, even in the case of $Q_{c} = 10$, a white dwarf that possesses a magnetic field of $10^9 \rm G$ will require a radius of $\approx 0.6 R_{\odot}$ to support coherent radio emission at a period of 6.45 hours.

\subsection*{Neutron Star}
The spin-down luminosity of \ulpo{} can be calculated by
\begin{equation}
    \dot{E} = 4\pi I \dot{P} P^{-3} \lesssim 10^{26} \luminu,
\end{equation}
where $I \sim 10^{45} \, \rm g\, cm^{-2}$ is the moment of inertia of a typical neutron star. This is an order smaller than the isotropic radio luminosity of \ulpo{}, suggesting that the radio pulses are not powered by a spin-down mechanism.

The pulse narrowing with frequency seen in the MeerKAT observation is roughly in concordance with standard neutron star pulsar radius-to-frequency mapping where  $W \propto \nu^{-\alpha}, \, \alpha \sim -0.6$ to $-0.3$ \citep{Thorsett1991,Kijak_Gil_2003}. The phase-dependent polarization, narrow duty cycle, and narrowing of pulse width with frequency suggest the emission emerges or decouples (at different heights) from a magnetically-dominated plasma where the local field bundle tangent is the observer viewing direction, well within the light cylinder of the rotator. This requires a magnetized compact object and a power source to sustain strong particle acceleration, and possibly pair production.

Therefore, a magnetar is more likely to be the progenitor of \ulpo{}.  The decaying radio luminosity of \ulpo{} is also consistent with that observed in the six currently known radio-loud magnetars whose radio emission is triggered after an outburst and then fades away \citep{MagnetarReview}. However, \ulpo{} has a much steeper spectrum compared to magnetars, which have $\alpha \gtrsim -0.5$ \citep{MagnetarReview}. Although some magnetars exhibit steeper or varying spectral indices \citep{XTE1810,PSRJ1818_repeated}, this remains an unusual characteristic. Additionally, none of these magnetars exhibit an interpulse, a characteristic that could simply be due to unfavorable viewing geometries. The presence of an interpulse in \ulpo{} is a unique feature that positions it distinctively to provide insights into the structure of the emission beam and the associated viewing geometry.

\section*{Optical and Near-Infrared searches}
We identified archival data covering the position of \ulpo{}. The source was covered by the Panoramic Survey Telescope and Rapid Response System (Pan-STARRS; \citep{ps1}) $3\pi$ survey in the $grizy$ bands, as well as by the VST Photometric H$\alpha$ Survey of the Southern Galactic Plane and Bulge (VPHAS+; \citep{vphas}) in the $ugri$ bands. For Pan-STARRS, we downloaded the coadded data from their archive and determined updated photometric solutions relative to the PS1 data. For VPHAS+ we downloaded individual exposures from the European Southern Observatory (ESO) archive and examined the co-added VPHAS+ images. This included $3\times 40\,$s in the $g$ filter, $4\times 25\,$s in the $r$ filter, and $2\times 25\,$s in the $i$ filter. The individual exposures were coadded using \texttt{swarp} \citep{swarp}.  We determined 3$\sigma$ upper limits of $g=22.7$, $r=22.9$, and $i=21.8$ (AB): these were reasonable given the pipeline-produced upper limits for the individual exposures.  The limiting magnitudes in this region from the PS1 stacked catalogs are $g=23.7$, $r=23.6$, $i=23.5$, and $z=22.9$ (AB).

In addition to these archival observations, we obtained deeper near-infrared observations of our own.  We observed \ulpo{} in the near-infrared (NIR) $J$-band with the Wide-field Infrared Camera (WIRC; \citep{WIRC}) on the Palomar 200-inch (Hale) telescope. The observations comprised  $9\times 45\,$s dithered exposures, corresponding to total exposure time of 405\,s. The data were dark subtracted and flat-fielded using python functions and were  astrometrically and photometrically calibrated by comparing to nearby stars from the 2MASS Point Source Catalog \citep{2MASSPSC}.
We then observed \ulpo{} on 2024-July-18 for 30\,min using the near-infrared ($2.2\,\mu$m) $K_s$ filter of the FourStar camera \citep{FourStar} on the Magellan Baade 6.5-m telescope. Data were reduced using FourCLift \citep{FourClift}, which corrects for dark current and non-linearity and subtracts the sky background using a bivariate wavelet model. A total of $300 \times 5.8\,$s frames were stacked accounting for camera distortion and variations in the sky background. The images were photometrically calibrated using unsaturated 2MASS \citep{2mass} stars. Seeing was $\approx 0.8^{\prime\prime}$. 

\section*{X-ray searches}
\subsection*{Swift}\label{sec:swift}
Based on the measured DM of $188\, \dmunits$, we assume an X-ray column density of $N_H \approx 6 \times 10^{21}\, \rm cm^{-2}$ \citep{XrayColumnDensity}. Following \citep{ConstrainingGLEAMX}, we calculated the unabsorbed flux for two spectral models: a power-law with photon index $\rm \Gamma = 2$, which assumes non-thermal emission from a pulsar/magnetar, and a blackbody with $kT = 0.3\, \rm keV$, which represents thermal emission from a pulsar/magnetar. The processing was done using \texttt{PIMMS}, where we assumed response from the XRT detector and PC filter. Based on a count rate of $4.55 \times 10^{-3} \, \rm counts/s$, we infer an unabsorbed flux limit of $F_{BB} < 2.44 \times 10^{-13}\, \Xrayunits$ for the blackbody model, and $F_{PL} < 3.87\, \times 10^{-13} \Xrayunits$ for the power-law model. This implies an upper limit in isotropic X-ray luminosity of $L_X \approx 7.4 \times 10^{32}\, \luminu$.

\subsection*{NICER}\label{sec:nicer}
The source was also observed with \textit{NICER} to look for any contemporaneous  X-ray flaring with the radio emission. We first generated level-2 data products using \textsc{NICERDAS} as part of the \textsc{HEASOFT} suite~\footnote{\url{https://heasarc.gsfc.nasa.gov/docs/software/heasoft/}}. The resulting lightcurves were checked for strong particle flaring at the harder X-ray energies. These flares were visually identified and removed to create Good Time Intervals (GTI) for the entire NICER dataset. We did not detect any X-ray enhancement during these observations. In order to compute an upper-limit on the X-ray flux of any X-ray flare, we first computed the background lightcurve for all the observations using the \texttt{SCORPEON} background model \footnote{\url{https://heasarc.gsfc.nasa.gov/docs/nicer/analysis_threads/scorpeon-overview/}}. We estimated a mean background rate of $0.55 \, \rm counts~s^{-1}$. Assuming a 5-$\sigma$ limit, for a standard deviation of 0.12~counts~$\rm s^{-1}$ we estimated that we would need a peak count rate of 1.16~counts~$\rm s^{-1}$. Then, we converted this upper-limit on the count rate to a peak X-ray flux using the \texttt{WebPIMMS} interface~\footnote{\url{https://heasarc.gsfc.nasa.gov/cgi-bin/Tools/w3pimms/w3pimms.pl}}. Assuming that the X-ray pulses/flares are thermal in nature with a black-body temperature of 0.3~keV and using an N$_H$ values of $\approx 5.6 \times 10^{21}\, \rm cm^{-2}$ , we report an upper-limit on the 0.2-12~keV peak X-ray flux of 1.6$\times 10^{-12}$~ergs~cm$^{-2}$~s$^{-1}$. This corresponds to an X-ray luminosity upper-limit of $\approx$3.0$\times$10$^{33}$~\luminu.

\section*{Archival Radio searches}
\subsection*{VLA}\label{sec:vla_archive}
We obtained the archival observations from the Karl G. Jansky Very Large Array (VLA) data archive\footnote{VLA archival observations can be found \href{here}{https://data.nrao.edu/portal/#/}.}. We looked for observations that had \ulpo{} within the primary beam of the telescope which resulted in five different observations under the project code: 13A$-$120. Each observation was two minutes in duration and were conducted at L-band (1--2\,GHz) with a time resolution of 5\,seconds. The raw data was calibrated using 3C286 as the bandpass calibrator and J1822$-$0938 as the phase calibrator. Calibrated observations were then cleaned and deconvolved using the \texttt{tclean} routine from \texttt{CASA} to generate a background model. This was followed by a sky model subtraction to obtain the background-subtracted visibilities. We excised baselines shorter than 250\,m to remove any diffuse background that might contribute to the source flux. We then estimated the source fluxes by fitting for a point source at the source location. 

No point source was detected in the time-integrated images. Hence, we phase-rotated the observations to the \ulpo{} location and generated the dynamic spectra for the target by averaging all the baselines. We examined the dynamic spectra and searched for fainter burst-like emission. We then estimated the upper limit on the burst-like emission by quoting the noise from the time-resolved light curve at 5-second resolution. Since the scan duration is less than the burst duration seen elsewhere, no intermediate averaging was done.

\subsection*{ASKAP}
We obtained all publicly available archival ASKAP observations that covered the field of \ulpo{} with the source within the full width at half maximum of the antenna. This resulted in various observations made under the project codes AS207 (VAST), AS110 (RACS), and AS113 (ULP1). Calibrated data were phase-rotated to the respective beam centers cleaned and model-subtracted using similar techniques described in Section \ref{sec:vla_archive}. Full-time integrated images and dynamic spectra were also generated similarly. VAST, RACS, and ULP1 observations lasted for 12 minutes, 15 minutes, and 60 minutes, respectively. Given the pulse widths, we also searched for the existence of fainter bursts, in ULP1 observations, by rebinning the light curve at 10-minute intervals, which is roughly the pulse width of \ulpo{}. We did not detect any point source or significant bursts at lower flux density levels in either the raw light curve with 10-second resolution or the rebinned light curve.

\backmatter


\section*{Declarations}


\begin{itemize}

\item \textbf{Funding}
T.M., Y.W.J.L., J.N.J.S., and R.M.S. acknowledge funding from the Australian Research Council Discovery Project DP\,220102305. M.C. acknowledges support of an Australian Research Council Discovery Early Career Research Award (project number DE220100819) funded by the Australian Government. Parts of this research were conducted by the Australian Research Council Centre of Excellence for Gravitational Wave Discovery (OzGrav), project number CE230100016. R.M.S. and N.H.W. acknowledge support through Australian Research Council Future Fellowships FT190100155, and FT190100231, respectively. M.G. and C.W.J. acknowledge support through the Australian Research Council's Discovery Projects funding scheme (DP210102103). Z.~W. acknowledges support by NASA under award number 80GSFC21M0002.
The development of the CRACO system has been supported through Australian Research Council Linkage Infrastructure Equipment and Facilities grant LE210100107.

\item \textbf{Data availability} 
The data that support the findings of this study are available at Zenodo: \url{https://doi.org/10.5281/zenodo.14043008}
All ASKAP data are publicly available via CASDA (\url{https://research.csiro.au/casda/}). 

\item \textbf{Code availability} The timing was performed using \textsc{tempo2} \citep{tempo2} and \textsc{PINT} \citep{PINT}. Specific Python scripts used in the data analysis are available on request from Y.W.J.L. and M.C.

\item \textbf{Acknowledgements} 
We would like acknowledge Matthew Bailes and Laura Spitler as co-PIs of the CRACO LIEF grant LE210100107. 
We would like to thank Matthew Bailes for supporting the PTUSE backend machine used in the MeerKAT observation. We would also like to thank Zaven Arzoumanian and the NICER team for their assistance in conducting X-ray observations.  We are grateful to the ASKAP engineering and operations team for their assistance in supporting the observations. This scientific work uses data obtained from Inyarrimanha Ilgari Bundara / the Murchison Radio-astronomy Observatory. We acknowledge the Wajarri Yamaji People as the Traditional Owners and native title holders of the Observatory site. CSIRO’s ASKAP radio telescope is part of the Australia Telescope National Facility (\url{https://ror.org/05qajvd42}). Operation of ASKAP is funded by the Australian Government with support from the National Collaborative Research Infrastructure Strategy. ASKAP uses the resources of the Pawsey Supercomputing Research Centre. Establishment of ASKAP, Inyarrimanha Ilgari Bundara, the CSIRO Murchison Radio-astronomy Observatory and the Pawsey Supercomputing Research Centre are initiatives of the Australian Government, with support from the Government of Western Australia and the Science and Industry Endowment Fund.

This manuscript makes use of data from MeerKAT (Project ID: DDT-20240209-JL-01) and ATCA (Project ID: C3363). We would like to thank SARAO for the approval of the MeerKAT DDT request and the science operations and CAM/CBF and operator teams for their time and effort invested in the observations. The MeerKAT telescope is operated by the South African Radio Astronomy Observatory, which is a facility of the National Research Foundation, an agency of the Department of Science and Innovation (DSI). 
This scientific work uses data obtained from telescopes within the Australia Telescope National Facility $\footnote{\url{https://ror.org/05qajvd42}}$ which is funded by the Australian Government for operation as a National Facility managed by CSIRO. PTUSE was developed with support from the Australian SKA Office and Swinburne University of Technology. This work has made use of the NASA Astrophysics Data System.

\item \textbf{Author contributions}
Y.W.J.L. and M.C. drafted the manuscript with suggestions from co-authors, and are the PIs of the MeerKAT data. Y.W.J.L. reduced and analysed the MeerKAT imaging data and the ATCA data, analysed the ASKAP data, and performed astrometry on the source. M.C. reduced the MeerKAT PTUSE data with Y.W.J.L. E.L. calibrated and reduced the ASKAP data. D.L.K. and S.M. conducted pulsar timing on the source. D.L.K. performed the Pan-STARRS and VPHAS+ archive search. T.M., L.F., Z.W., and N.H.W. contributed to discussions about the nature and emission mechanism of the source. A.A. performed the ASKAP and VLA archival searches and analyses. N.H.W. reduced the MWA data and analysed the spectral index. V.K. and M.M.K. performed the WIRC observation and calibrated the data. S.O. performed the FourStar camera observation and calibrated the data. H.Q. performed the Swift observation and analysed the data. K.M.R. and K.G. performed the NICER observation and analysed the data. A.Z. and M.E.L. analysed the rotation measure of the source. K.W.B., A.D., C.J., and R.M.S. are the PIs of CRACO. M.G., V.G., J.N.J.S., A.J., Y.W.J.L., P.U., Y.W., and Z.W. are the builders of CRACO. T.M. is the PI of the ATCA project C3363. T.M. and D.L.K. and the PIs of VAST, and D.D. and L.D. are the Project Scientists of VAST. The PIs and builders of VAST and CRACO coordinated the initial investigation of \ulpo{}.

\item \textbf{Conflict of interest/Competing interests} 
The authors declare no competing interests.

\end{itemize}

\end{document}